\documentclass[12pt]{article}
\usepackage{graphicx,amsmath,amssymb}
\begin{document}
%%%%%%%%%%%%%%%%%%%%%%%%%%%%%%%%%%%%%%%%%%%

\def\a{\alpha}
\def\b{\beta}
\def\c{\varepsilon}
\def\d{\delta}
\def\e{\epsilon}
\def\f{\phi}
\def\g{\gamma}
\def\h{\theta}
\def\k{\kappa}
\def\l{\lambda}
\def\m{\mu}
\def\n{\nu}
\def\p{\psi}
\def\q{\partial}
\def\r{\rho}
\def\s{\sigma}
\def\t{\tau}
\def\u{\upsilon}
\def\v{\varphi}
\def\w{\omega}
\def\x{\xi}
\def\y{\eta}
\def\z{\zeta}
\def\D{\Delta}
\def\G{\Gamma}
\def\H{\Theta}
\def\L{\Lambda}
\def\F{\Phi}
\def\P{\Psi}
\def\S{\Sigma}

\def\o{\over}
\def\beq{\begin{eqnarray}}
\def\eeq{\end{eqnarray}}
\newcommand{\gsim}{ \mathop{}_{\textstyle \sim}^{\textstyle >} }
\newcommand{\lsim}{ \mathop{}_{\textstyle \sim}^{\textstyle <} }
\newcommand{\VEV}[1]{ \left\langle {#1} \right\rangle }
\newcommand{\bra}[1]{ \langle {#1} | }
\newcommand{\ket}[1]{ | {#1} \rangle }
\newcommand{\EV}{ {\rm eV} }
\newcommand{\KEV}{ {\rm keV} }
\newcommand{\MEV}{ {\rm MeV} }
\newcommand{\GEV}{ {\rm GeV} }
\newcommand{\TEV}{ {\rm TeV} }
\def\diag{\mathop{\rm diag}\nolimits}
\def\Spin{\mathop{\rm Spin}}
\def\SO{\mathop{\rm SO}}
\def\O{\mathop{\rm O}}
\def\SU{\mathop{\rm SU}}
\def\U{\mathop{\rm U}}
\def\Sp{\mathop{\rm Sp}}
\def\SL{\mathop{\rm SL}}
\def\tr{\mathop{\rm tr}}

\def\IJMP{Int.~J.~Mod.~Phys. }
\def\MPL{Mod.~Phys.~Lett. }
\def\NP{Nucl.~Phys. }
\def\PL{Phys.~Lett. }
\def\PR{Phys.~Rev. }
\def\PRL{Phys.~Rev.~Lett. }
\def\PTP{Prog.~Theor.~Phys. }
\def\ZP{Z.~Phys. }

%%%%%%% added by Fumi %%%%%%%%%%
% FROM HERE
%\newcommand{\beq}{\begin{equation}}   
%\newcommand{\eeq}{\end{equation}}
\newcommand{\bea}{\begin{eqnarray}}   
\newcommand{\eea}{\end{eqnarray}}
\newcommand{\bear}{\begin{array}}  
\newcommand {\eear}{\end{array}}
\newcommand{\bef}{\begin{figure}}  
\newcommand {\eef}{\end{figure}}
\newcommand{\bec}{\begin{center}}  
\newcommand {\eec}{\end{center}}
\newcommand{\non}{\nonumber}  
\newcommand {\eqn}[1]{\beq {#1}\eeq}
\newcommand{\la}{\left\langle}  
\newcommand{\ra}{\right\rangle}
\newcommand{\ds}{\displaystyle}
\def\SEC#1{Sec.~\ref{#1}}
\def\FIG#1{Fig.~\ref{#1}}
\def\EQ#1{Eq.~(\ref{#1})}
\def\EQS#1{Eqs.~(\ref{#1})}
\def\REF#1{(\ref{#1})}
\def\GEV#1{10^{#1}{\rm\,GeV}}
\def\MEV#1{10^{#1}{\rm\,MeV}}
\def\KEV#1{10^{#1}{\rm\,keV}}
\def\lrf#1#2{ \left(\frac{#1}{#2}\right)}
\def\lrfp#1#2#3{ \left(\frac{#1}{#2} \right)^{#3}}
\def\oten#1{ {\mathcal O}(10^{#1})}
\def\oone{ {\mathcal O}(1)}
\def\oTEV#1{{\cal O}(10^{#1}){\rm\,TeV}}
\def\oGEV#1{{\cal O}(10^{#1}){\rm\,GeV}}
\def\oMEV#1{{\cal O}(10^{#1}){\rm\,MeV}}
\def\okEV#1{{\cal O}(10^{#1}){\rm\,keV}}
% UNTIL HERE

%%%%%%%%%%%%%%%%%%%%%%%%%%%%%%%%%%%%%%%%%%%%%%%%%%%%%%%%%%%%%%%%%%%%

\baselineskip 0.7cm

\begin{titlepage}

\begin{flushright}
UT-11-30\\
TU-889\\
IPMU 11-0146
 \end{flushright}

\vskip 1.35cm
\begin{center}
{\large \bf 
On the Adiabatic Solution to the Polonyi/Moduli Problem
}
\vskip 1.2cm
Kazunori Nakayama$^a$,
Fuminobu Takahashi$^{b,c}$
and 
Tsutomu T. Yanagida$^{a,c}$

\vskip 0.4cm

{\it $^a$Department of Physics, University of Tokyo, Tokyo 113-0033, Japan}\\
{\it $^b$Department of Physics, Tohoku University, Sendai 980-8578, Japan}\\
{\it $^c$Institute for the Physics and Mathematics of the Universe,
University of Tokyo, Kashiwa 277-8568, Japan}

\vskip 1.5cm

\abstract{
One of the solutions to the cosmological Polonyi problem is to introduce a large coupling between
the Polonyi field and the inflaton so that the Polonyi field adiabatically tracks the temporal minimum of the potential.
We study general conditions for the adiabatic suppression mechanism to work,
and find that a non-negligible amount of  the Polonyi field 
is induced in the form of coherent oscillations at the end of inflation.
In the case of low reheating temperature, this contribution is so small that it does not cause cosmological problems.
On the other hand, this contribution may be significant for a relatively high reheating temperature and we still need 
some amount of tuning in order to avoid the Polonyi problem.
We also point out that Polonyi particles produced from thermal plasma pose a severe constraint on the reheating temperature. 
Furthermore, we extend the original framework to include enhanced couplings of
the Polonyi field with the visible particles as well as with itself, and derive upper bounds on the
reheating temperature after inflation.
We also investigate the adiabatic solution to the cosmological moduli problem 
in gauge and anomaly mediation.
}
\end{center}
\end{titlepage}

\setcounter{page}{2}

%%%%%%%%%%%%%%%%%%%%%%%%%%%%%%%%%%%%
\section{Introduction}
%%%%%%%%%%%%%%%%%%%%%%%%%%%%%%%%%%%%

Supersymmetric (SUSY) extension of the standard model (SSM)
is a plausible candidate for physics beyond the standard model
since it solves the gauge hierarchy problem and leads to the successful gauge coupling unification at the
grand unification scale. It also contains candidates for dark matter (DM).

The gravity-mediation models for SUSY breaking are attractive because of its simplicity:
the SSM gaugino masses arise from an $F$-term of an elementary gauge singlet field $Z$
through Planck-scale suppressed interactions. However, the Polonyi field $Z$ is known to cause
a serious cosmological problem.
Since it is neutral under any symmetry,  there is no special point
in its field space. Therefore,  the minimum of the effective potential for the Polonyi field during inflation is
generically deviated from the one in the low energy.
After inflation, the Polonyi starts to oscillate about the minimum with an amplitude of
order the Planck scale $M_P$, and soon dominates the energy density of
the Universe.  Since its interactions are suppressed by the Planck scale,
the  lifetime of the Polonyi is very long, leading to an onset of a
radiation-dominated Universe with a low temperature, typically below
MeV. Such a low temperature would dramatically alter the big bang
nucleosynthesis (BBN) predictions of light element abundances in
contradiction with observations.  This is the notorious cosmological Polonyi
problem~\cite{Coughlan:1983ci}.

Even if one gives up the gravity mediation, 
there may be  scalars having similar properties to the Polonyi field. Indeed,
there generally appear such scalars, called moduli, associated with
 the compactification of extra dimensions in the string theory.
The moduli have a cosmology similar to the Polonyi, and cause a  serious cosmological 
problem~\cite{Banks:1993en}.

Some of the solutions to the cosmological Polonyi/moduli problem require significant modification
of the conventional cosmological scenarios.
One of them is to make the Polonyi/moduli heavy enough to decay well before
the BBN  begins.
In the anomaly-mediated SUSY breaking models~\cite{Randall:1998uk}, 
the gravitino as well as the moduli are expected to have masses much heavier than the weak scale.
However, the solution turned out to be more complicated, because of unsuppressed production
of SUSY particles including gravitinos  from the modulus decay~\cite{Endo:2006zj}.
The modulus decay also significantly dilutes the pre-existing baryon asymmetry. 
This makes the most baryogenesis mechanisms unsuccessful, although it is still possible to create 
the right amount of baryon asymmetry by the Affleck-Dine mechanism~\cite{Affleck:1984fy,Kawasaki:2007yy}.
Another solution is to introduce late-time entropy production 
for diluting the Polonyi/moduli abundance. For instance, the thermal inflation~\cite{Yamamoto:1985rd} can provide
enough dilution.
Again, it also dilutes any pre-existing baryon asymmetry to a negligible amount.
Considering that thermal inflation should dilute the modulus abundance by up to a factor of $\sim 10^{20}$
in order to solve the moduli problem, it is hopeless to try to create a sufficient amount of  baryon asymmetry 
 which survives the huge dilution.
Thus we need to create the baryon asymmetry after thermal inflation by an elaborate mechanism~\cite{Stewart:1996ai}.

An interesting solution to the Polonyi/moduli problem was proposed long ago by Linde~\cite{Linde:1996cx}.
It was pointed out that, if the modulus field has a mass squared of $\sim c^2 H^2$ before it starts to oscillate,
with $c= {\cal O}(10)$ and $H$ being the Hubble parameter,
the modulus follows the time-dependent potential minimum adiabatically and 
the resultant amplitude of coherent oscillations is significantly suppressed.
This simple solution is attractive  since it works for a wide range of the Polonyi/modulus mass, 
and  since there is no need for an additional late-time entropy production, 
which may make the standard leptogenesis scenario~\cite{Fukugita:1986hr} viable.
Recently, two of the present authors (FT and TTY) noticed that 
there is an upper bound on the reheating temperature for the adiabatic solution to work~\cite{Takahashi:2011as}
and also showed that such a large Hubble mass may be a consequence of the 
strong dynamics at the Planck scale~\cite{Takahashi:2010uw} or the fundamental cut-off scale
 one order of magnitude lower than the Planck scale~\cite{Takahashi:2011as}.

In this paper we study the adiabatic suppression mechanism  in a great detail
in order to establish the solution in a complete form and explore parameter space where
the Polonyi/moduli problem is solved.
First, we carefully investigate the issue of adiabaticity and how the Polonyi/moduli oscillation
is induced after inflation.
We find that the coherent oscillations of the Polonyi/modulus field are generically induced at the end of inflation
where the adiabaticity is violated and the produced amount  depends on inflation models.
The adiabatic solution may not work without tuning parameters or an additional entropy production 
for a relatively high reheating temperature, 
although the required amount of tuning or entropy production is greatly relaxed with respect to the original Polonyi/moduli problem.
For a low reheating temperature, on the other hand, the adiabatic solution can solve the Polonyi/moduli problem
without any fine-tuning.

Next, we point out that, even if coherent oscillations of the Polonyi/moduli 
are reduced to a negligible amount by the adiabatic solution (and some fine-tuning), 
the Polonyi/moduli are generically produced by particle scatterings
in thermal plasma as long as they have (Planck-suppressed) interactions with the
SSM particles. In particular, the thermal production of the Polonyi field is inevitable because it must be coupled
to the visible sector to mediate the SUSY breaking.
We find that the thermal production of the Polonyi/moduli fields leads to non-trivial cosmological constraints.
In the gravity mediation, the constraints are so stringent that 
there is no parameter space where the thermal leptogenesis scenario works, 
other than the heavy gravitino mass region, $m_{3/2} \gtrsim 10$\,TeV. 
 
Then, based on the findings of Refs.~\cite{Takahashi:2010uw,Takahashi:2011as},
we extend the original framework to include an enhanced coupling of the Polonyi field with itself
as well as the SSM particles. 
We will see that a new interesting possibility emerges in this case,
where the gravitino is the lightest SUSY particle (LSP) and relatively high-reheating temperature is allowed.
The focus-point like mass spectrum is favored in this case.

We will also discuss the implications of the adiabatic solution to the
moduli problem in the gauge-mediated SUSY breaking models where the moduli and gravitino are light.
For light moduli, $m_{z} \lesssim 1\,$MeV, the adiabatic suppression mechanism can solve the moduli problem
without fine-tuning on the modulus potential. 
However, we show that the reheating temperature is severely bounded from above
due to the gravitino thermal production and the condition for the adiabaticity,
and the thermal/non-thermal leptogenesis does not work even in a scenario with an extremely
light gravitino ($m_{3/2} \lesssim 16$~eV~\cite{Viel:2005qj}).\footnote{
	The extremely light modulus with a mass smaller than $\mathcal O(1)$~keV in the form of coherent oscillations 
	contributes to the DM density of the Universe.
	Hence anthropic arguments may guarantee the smallness of the modulus abundance.
}

Lastly we will discuss the adiabatic solution to the
moduli problem in the anomaly-mediated SUSY breaking models where the moduli and gravitino are heavy
enough to decay before BBN.
In this case, the adiabatic suppression can solve the moduli problem for a relatively high reheating temperature
once we allow a tuning of $\mathcal O(0.01)$ on the modulus potential.
The (non-)thermal leptogenesis may work for the gravitino mass of $\sim 100$\,TeV
with the Wino LSP with mass of a few hundred GeV.

The rest of this paper is organized as follows. 
In Sec.~\ref{sec:2} we investigate the adiabatic suppression mechanism,
and estimate the abundance of the Polonyi/moduli field in the form of coherent oscillations. We also
derive an upper bound on the reheating temperature for the mechanism to work~\cite{Takahashi:2011as}.
In Sec.~\ref{sec:2-5} we study the Polonyi/moduli oscillation induced at the end of inflation in detail
and show that it significantly contributes to the final Polonyi/moduli abundance.
We derive the required amount of fine-tuning to solve the moduli problem.
In Sec.~\ref{sec:3} we study the Polonyi problem in the gravity mediation in detail in order to clarify to what extent the Polonyi abundance is suppressed
in the mechanism, taking account of thermal production of the Polonyi field. In Sec.~\ref{sec:4} and \ref{sec:5} we will discuss the moduli
problem in gauge and anomaly mediation. The last section is devoted to the discussion and conclusions.

%%%%%%%%%%%%%%%%%%%%%%%%%%%%%%%%%%%%
\section{Adiabatic solution to the Polonyi/moduli problem}
\label{sec:2}
%%%%%%%%%%%%%%%%%%%%%%%%%%%%%%%%%%%%

%%%%%%%%%%%%%%%%%%%%%%%%%%%%%%%%%%%%
\subsection{Basic idea}
\label{2-1}
%%%%%%%%%%%%%%%%%%%%%%%%%%%%%%%%%%%%

First we briefly review the basic idea to suppress the modulus abundance~\cite{Linde:1996cx}.
Let $z$  denote collectively a modulus field, including the Polonyi field.
We set the origin of $z$ so that it 
coincides with the potential minimum at present. In the early Universe, the
effective potential of $z$ could receive various corrections from its interactions with
the inflaton and/or SSM particles.  In particular, if $z$ has a quartic coupling 
with the inflaton in the K\"ahler potential, $z$ receives the  so-called Hubble-induced
mass term. Here and in what follows, the inflaton also refers to a field which dominates
the energy density of the Universe when the modulus starts oscillating.

Let us express the effective potential in the early Universe as
\begin{equation}
	V = \frac{1}{2}m_z^2 z^2 + \frac{1}{2}c^2H^2(z-z_*)^2,   \label{Vmod}
\end{equation}
where $m_z$ is the mass of $z$ in the low energy,
$z_*$ is the initial displacement during inflation, $H$ is the Hubble parameter and $c$ is a constant.
The amplitude $z_*$ is expected to be of order the Planck scale.
For simplicity we treat $z$ as a real scalar field  when we consider its dynamics, but this does not
affect the following argument. We have assumed that the potential can be approximated
by a quadratic potential about the origin, at least up to $z= z_*$. 
This is expected to be the case for the Polonyi field, but
the potential may take a more general form. We will come back to this issue in Sec.~\ref{sec:2-5}.

Let us first consider the case of $c = \mathcal O(1)$. This is the case if the modulus and the inflaton are coupled by
the Planck-suppressed operators with coefficients of order unity.
The modulus dynamics is as follows. 
Assuming that the Hubble parameter during inflation is much larger than $m_z$, 
$z$  is stabilized at $z \simeq z_*$ during inflation.
After inflation, it begins to oscillate around the minimum $z=0$
with an amplitude of $z_*$ when $H\sim m_z$. The modulus abundance in the form of coherent oscillations is given by
\begin{equation}
	\frac{\rho_z}{s} = \left \{ \begin{array}{ll}
	 \displaystyle
	  \frac{1}{8}T_{\rm R} \left( \frac{z_*}{M_P} \right)^2 & ~~~{\rm for~~}\Gamma_\phi < m_z \\
	  \displaystyle
	 \frac{1}{8}T_{\rm osc} \left( \frac{z_*}{M_P} \right)^2& ~~~{\rm for~~}\Gamma_\phi > m_z
	\end{array}
	\right. ,   \label{analytic1}
\end{equation}
where $\rho_z$ is the modulus energy density, $s$ is the entropy density,
$\Gamma_\Phi$ is the inflaton decay rate,
$T_{\rm R} \equiv (10/\pi^2 g_*)^{1/4}\sqrt{\Gamma_\phi M_P}$ denotes the reheating temperature after inflation, 
and $T_{\rm osc} \equiv (10/\pi^2 g_*)^{1/4}\sqrt{m_z M_P}$.
The above modulus abundance is so large that it causes a serious cosmological problem.
For example, we need a tuning of $z_* \lesssim 10^{-10}M_P$ to satisfy the BBN bound
for a typical reheating temperature $T_{\rm R}=10^6$\,GeV and $m_z\sim 1$\,TeV. 
If $c$ is much smaller than $\mathcal O(1)$, the modulus abundance depends on the initial displacement,
which is subject to quantum fluctuations during inflation. 
As long as it is of order $z_*$, the resultant modulus abundance is the same order as
in the case of $c= \mathcal O(1)$.

The situation significantly changes if $c \gg \mathcal O(1)$.
One might expect that the modulus begins to oscillate when $H \sim c^{-1} m_z$ with an amplitude of $z_*$,
when the potential minimum starts to move from $z_*$ to the origin.
However, this  is not true.
At $H\sim c^{-1} m_z$, the Hubble parameter is much smaller than $m_z$,  which 
means that the potential minimum moves more slowly than the 
typical time scale for the modulus dynamics. 
Thus there is enough time for $z$ to follow  the potential minimum, and
as a result, the coherent oscillations of $z$ are significantly suppressed.

\begin{figure}[htbp]
\begin{center}
\includegraphics[scale=1.5]{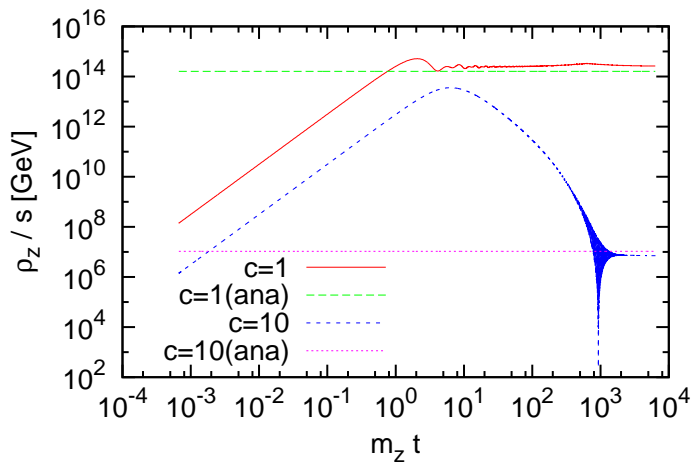}
\caption{
	Time evolution of the modulus abundance for $c=1$ and $c=10$.
	For comparison, analytic estimations, labeled as (ana), are also plotted.
	We have taken $\Gamma_\phi=10^{-2.5} m_z$ and $z_* = M_P/c$. 
 }
\label{fig:model1}
\end{center}
\end{figure}

It is possible to estimate the suppression factor analytically, by solving the equation of motion for the modulus
in the potential (\ref{Vmod}).
The final amplitude of the coherent oscillations is suppressed by the following
factor~\cite{Linde:1996cx}
\beq
{\cal S} \;=\; \frac{3\sqrt{2 p \pi }}{4} c^{(3p+1)/2}\, \exp{\left(-\frac{c p \pi}{2}\right)},
\label{supfac}
\eeq
where $p$ parametrizes the Hubble parameter as $H = p/t$ when $H=c^{-1} m_z$, and it is
given by $p=2/3$ and $p=1/2$ before and after the reheating, respectively.
Thus, as we increase $c$ for fixed $m_z$ and $z_*$, the final modulus abundance is 
exponentially suppressed by ${\cal S}^2$ relative to the estimate (\ref{analytic1}).
In Fig.~\ref{fig:model1} we have shown the numerically obtained 
evolution of the modulus abundance as a function of time
as well as the analytic estimates based on
(\ref{analytic1}) and (\ref{supfac}) for $c = 1$ (top) and $10$ (bottom).
Here we took $z_* = M_P/c$.
We can see that in both cases the analytic estimates agree well with the numerical results. 
In this plot, the entropy density $s$ is defined as $s(t)=s(T_{\rm R})[  a(T_{\rm R})/a(t) ]^3$
and $s(T_{\rm R})= (2\pi^2 g_{*s}/45)T_{\rm R}^3$.
The definition of $s$, and hence $\rho_z/s$,  coincides with the standard one at late times after reheating.

%%%%%%%%%%%%%%%
%%%%%%%%%%%%%%%

Thus, we can achieve  sufficient suppression of the modulus
abundance for $c = O(10)$  because of the exponential factor appearing in Eq.~(\ref{supfac}), which reflects the fact that
 the variation of the adiabatic invariant is exponentially suppressed.
This is essentially what was pointed out in the pioneering paper \cite{Linde:1996cx}.

\vspace{5mm}

Lastly, we note here that the above estimate (\ref{supfac}) is based on the analysis of Ref.~\cite{Lyth:1995ka},
in which the initial condition is set to be $z= z_*$ and $\dot{z} = 0$ at $t=0$ when the Hubble parameter
is infinitely large. 
However, we would like to emphasize here that the actual initial condition of the modulus field is set during inflation when
the Hubble parameter is finite. 
In addition, the inflaton dynamics changes the modulus potential at the end of inflation with a time scale of the inflaton mass,
which may violate the adiabatic condition.
Therefore, the above analytic estimate is only approximate, and 
there is in general additional contribution to the modulus abundance produced at the end of inflation. 
That is to say, it is important to follow the modulus dynamics during the first oscillation $\Delta t = \mathcal O((cH)^{-1})$
after inflation, and non-negligible amount of the modulus abundance can be generated during the short period.
Once the modulus starts to oscillate, the particle production is indeed exponentially suppressed as explained above.
This additional particle production at the end of inflation has been overlooked so far in the context of the adiabatic suppression mechanism,
although similar kind of particle production was considered in the detailed analysis of thermal inflation~\cite{Lyth:1995ka}.
We will come back to this issue in Sec.~\ref{sec:2-5}, and derive the condition for the adiabatic suppression to work successfully.
For the moment, we will neglect the additional contribution.

%%%%%%%%%%%%%%%%%%%%%%%%%%%%%%%%%%%%
\subsection{The origin of the enhanced couplings}
%%%%%%%%%%%%%%%%%%%%%%%%%%%%%%%%%%%%

The essential ingredient for the mechanism described above is the large Hubble mass term
with $c \gg 1$. Suppose that the inflaton $\Phi$ has a quartic coupling with the modulus in the K\"ahler potential 
as~\cite{Linde:1996cx}
\beq
K \;\supset\; - c_1^2 \frac{ |\Phi|^2 |Z|^2 }{M_P^2},
\label{quartic}
\eeq
where $Z$ and $\Phi$ are chiral superfields corresponding to the modulus and the inflaton,
respectively. The $c$ is related to $c_1$ as
\begin{equation}
	c^2 = 3(c_1^2+1).
\end{equation}
If $c_1$ is larger than order unity for some reason, a large Hubble-induced mass
term for the modulus is generated. 
The origin of $c_1\gg1$ may be due to a strong dynamics at the Planck scale~\cite{Takahashi:2010uw}
or a cut-off scale one order of magnitude below the Planck scale~\cite{Takahashi:2011as}.

We would like to emphasize here that the Hubble-induced mass term generated by (\ref{quartic}) disappears after
the reheating, since the energy density of the Universe becomes dominated by radiation afterward. Therefore,
the reheating should occur sufficiently late for the adiabatic solution to work.
This places an interesting upper bound on the reheating temperature~\cite{Takahashi:2011as}, 
which we shall derive explicitly in the next subsection.

We note here some subtleties regarding the thermal effects. 
If the modulus is coupled to the SSM particles in plasma, the modulus potential
receives thermal corrections after the reheating.  In particular, if the thermal mass is much larger than the Hubble parameter,
the modulus adiabatically follows the time-dependent minimum of the potential
in a similar way discussed above.
The thermal mass is roughly estimated as
\begin{equation}
	%m_T^2 \sim \mathcal O(0.01)\times ({\rm number~of~particles~running~the~loop})\frac{T^4}{M^2},
	m_T^2 \sim c'^2 \frac{T^4}{M_P^2} = \mathcal O(0.01)\times  c'^2 H^2,
\end{equation}
where  we have used $H^2 = (\pi^2 g_*/90) T^4/M_P^2$ with $g_* \sim 230$ in the 
second equality. Here $c'$ is the coupling constant between the modulus and SSM
particles in thermal bath normalized by the Planck scale; it corresponds to e.g. $c_3$ or $c_5$ in Eqs.~(\ref{Zgg})
and (\ref{zzff}).
Thus, if $ c'$ is larger than $\mathcal O(100)$, the adiabatic solution may work even if the reheating
is completed before $H = m_z/c$. However, as we shall see later, such a large coupling between the modulus
and the SSM particles leads to problematic thermal production of the modulus. Therefore we assume
$c' \lesssim \mathcal O(10)$  in the following so that the thermal mass is at most comparable to the Hubble mass
with $c = \mathcal O(1)$
and it does not affect the modulus dynamics significantly.\footnote{
	There is also a linear term such as $V \sim T^4 z/M$~\cite{Buchmuller:2004xr}, 
	whose effects on the modulus abundance were discussed in Ref.~\cite{Nakayama:2008ks}.
}

%%%%%%%%%%%%%%%%%%%%%%%%%%%%%%%%%%%%
\subsection{Upper bound on the reheating temperature}
\label{2-3}
%%%%%%%%%%%%%%%%%%%%%%%%%%%%%%%%%%%%

Assuming that the large Hubble-induced mass term arises solely from an enhanced coupling between
the modulus and the inflaton, we can derive an upper bound on the reheating temperature 
for the solution to work. It is important to note that the large Hubble-induced mass term
disappears as $e^{-\Gamma_\phi t}$, where $\Gamma_\phi$ denotes the
decay rate of the inflaton. The temporal minimum moves on a time scale of $\Gamma_\phi^{-1}$ at the reheating,
which should be smaller than $m_z$ for the adiabatic solution to work.
Therefore, we have~\cite{Takahashi:2011as}
\beq
\Gamma_\phi \;\ll\; m_z.
\eeq
It depends on the required suppression factor as well as on the value of $c$
how much the decay rate should be suppressed compared to the modulus mass.
From Fig.~\ref{fig:cont}, one can see that the adiabatic solution does not work 
unless $\Gamma_\phi < 0.1 m_z$, and that the effect of the reheating becomes practically
negligible if $\Gamma_\phi < 0.01 m_z$. We therefore adopt $\Gamma_\phi < 0.05\, m_z$
as a reference value in the following analysis. 

We can express the bound in terms of the
reheating temperature $T_{\rm R}$ as~\cite{Takahashi:2011as}
\beq
T_{\rm R} \;\lsim\; 
%\lrfp{\pi^2 g_*}{10}{-\frac{1}{4}} \sqrt{0.05 m_z M_P}
%	\simeq 
3 \times \GEV{9} \lrfp{g_*}{230}{-\frac{1}{4}} \lrfp{m_z}{1{\rm\,TeV}}{\frac{1}{2}},
%2.8	
\label{eq:upperTR}
\eeq
where the reheating temperature $T_{\rm R}$ is related to the decay rate as
\beq
T_{\rm R} \;\equiv\; \lrfp{\pi^2 g_*}{10}{-\frac{1}{4}} \sqrt{\Gamma_\phi M_P}.    \label{TR_Gamma}
\eeq
The upper bound on the reheating temperature was overlooked so far, but it
has a very important implication especially for the baryogenesis scenario such as
thermal leptogenesis.

%%%%%%%%%%%%%%%%%%%%%%%%%%%%%%%%%%%%
\subsection{Suppressing the modulus abundance}
%%%%%%%%%%%%%%%%%%%%%%%%%%%%%%%%%%%%

We numerically evaluate the suppression factor for the modulus abundance, $\Delta$,
which is defined by the ratio of the actual modulus abundance to the 
analytic estimate for the case of $c=1$,
\begin{equation}
	\Delta \equiv \frac{\rho_z/s}{(\rho_z/s)_{\rm exp}},
\end{equation}
where the denominator is given by Eq.~(\ref{analytic1}).

\begin{figure}[htbp]
\begin{center}
\includegraphics[scale=0.7]{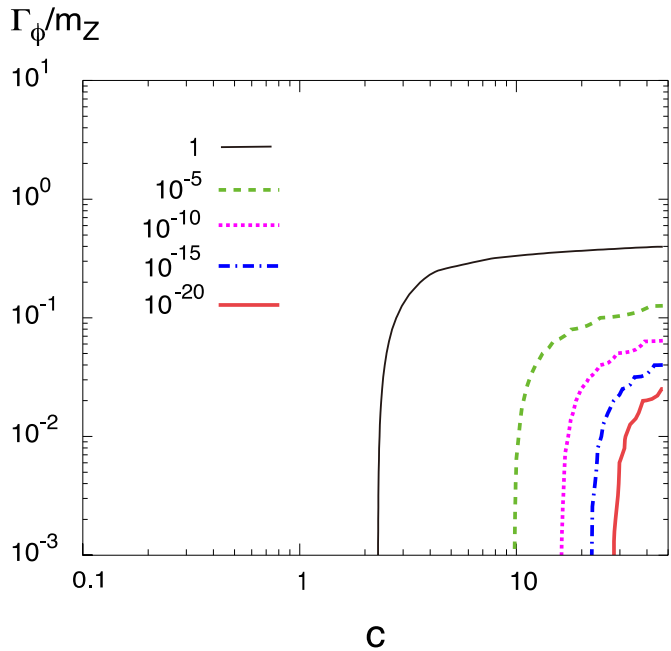}
\caption{
	Contours of the suppression factor $\Delta$ in the plane of $(c,\Gamma_\phi/m_z)$.
 }
\label{fig:cont}
\end{center}
\end{figure}

Contours of the suppression factor $\Delta$ in the plane of $(c,\Gamma_\phi/m_z)$ are shown 
in Fig.~\ref{fig:cont}.
The initial amplitude is taken to be $z_*=M_P/c$.\footnote{
	The large $c$ effectively corresponds to a small cutoff scale of order of $M_P/c$ in the 
	non-renormalizable K\"ahler potential.
	Thus the amplitude should be smaller than or comparable to this cutoff scale in order for the 
	effective description to be valid.
}
It is seen that the modulus abundance is highly suppressed for $c > 30$ and $\Gamma_\phi/m_z < 0.05$.
Practically, $\Delta \lesssim 10^{-20}$ is sufficient for satisfying the
BBN bound on the modulus abundance, that is, $\rho_z/s \lesssim 10^{-14}{\rm GeV}$ for $m_z \sim 1$~TeV
\cite{Kawasaki:2004qu}.
With this amount of suppression, the moduli do not dominate the Universe and therefore do not produce huge entropy with their decays.
Thus the baryon asymmetry is not diluted. 
On the other hand, if we merely demanded that the moduli do not cause entropy production,
the required suppression factor would be mild (say, $\Delta \lesssim 10^{-10}$ for $m_z =1$\,TeV and $T_{\rm R}\sim 10^6$\,GeV).
Hereafter we regard $\Delta \sim 10^{-20}$ as a typically required suppression factor for solving the 
cosmological moduli problem, although the precise constraint depends on the modulus mass and couplings, and also
the reheating temperature.

%%%%%%%%%%%%%%%
%%%%%%%%%%%%%%%

%%%%%%%%%%%%%%%%%%%%%%%%%%%%%%%%%%%%
\section{When does (not) the adiabatic solution work ?}
\label{sec:2-5}
%%%%%%%%%%%%%%%%%%%%%%%%%%%%%%%%%%%%

We have seen that the adiabatic solution works for $c = {\cal O}(10)$ in the potential (\ref{Vmod}).
However, there is in general an additional contribution to the Polonyi/modulus
abundance, and it is necessary to clarify under which conditions the adiabatic solution works.

Before going to concrete models, we
give a general discussion on the adiabatic suppression mechanism.
Suppose that the scalar potential for a modulus $z$ depends on the Hubble parameter.
During inflation the $z$ is stabilized at a point deviated  from the low-energy minimum, 
and the position of the minimum gradually changes as the Hubble parameter decreases after inflation.
Then question is in which situation the coherent oscillations are induced.
The key concept is the adiabaticity of the modulus dynamics in the time-dependent scalar potential.
Let us assume that a modulus with a mass larger than the Hubble parameter is oscillating
about the potential minimum $z_{\rm min}(t)$. 
If the rate of the change of the potential minimum is much smaller than the oscillation frequency,
the modulus number density in a comoving volume is conserved.
The condition is written as
\begin{equation}
	f(H)\equiv \left | \frac{\dot z_{\rm min}}{z_{\rm min}} \right | \ll 
	\left| \frac{V'(z_{\rm min} + \delta z)}{\delta z} \right|^{1/2},   \label{condition}
\end{equation}
where $\delta z$ denotes an amplitude of oscillations about the potential minimum.
If the potential can be approximated with a quadratic potential about $z_{\rm min}$,
the far right hand side becomes equal to the mass. 
If this condition is violated, particle production occurs as $z_{\rm min}$ moves.
Thus the condition (\ref{condition}) must be satisfied in order for the modulus amplitude to be suppressed.

A couple of comments are in order. First, the adiabatic suppression mechanism relies on the conservation of
the adiabatic invariant ($\approx$ the modulus number density in the comoving volume), 
which is defined for a periodic motion.  In other words, it does not say anything about the initial abundance
before the Polonyi/modulus field starts oscillating, which is determined by the
dynamics during the first period of oscillation. Therefore, careful case-by-case analysis is needed
in order to estimate the initial Polonyi/modulus abundance.

Secondly, if $\dot z_{\rm min}$ suddenly changes, the above condition \REF{condition} is
not sufficient. (We here assume that $z_{\rm min}$ does not jump
at the end of inflation.)
In this case, one should also consider a condition similar to \REF{condition} with  $f(H)$ replaced 
with the typical time scale of the dynamics which determines the evolution of the potential minimum. For instance, as we shall see below,
in a broad class of inflation models, the potential minimum  starts to move at the end of inflation in a time scale 
of the inflaton mass.  
In this case one should regard the inflaton mass as the typical time scale.
Then the adiabaticity condition is usually violated at the end of inflation, and coherent oscillations of
the Polonyi/modulus field are induced. We shall study this effect in \SEC{2-5-1}.

Another issue to be addressed is whether $z_{\rm min}$ remains same after inflation.
As we shall see in \SEC{2-5-2}, in realistic single-field inflation models in supregravity,  
because of the non-trivial couplings between the inflaton and 
the modulus in the supregravity potential, the position of the minimum generically changes in a non-adiabatic way, and
some amount of modulus oscillation is induced at the end of inflation.
We will also see in \SEC{2-5-3} that
in a class of multi-field inflation models, one (or more) of the fields acquires a non-zero $F$-term at the end of
inflation, which in general changes the position of the potential minimum. 
For instance, in the hybrid inflation, the waterfall field has a vanishing $F$-term during inflation, but it 
starts to oscillate after inflation, acquiring a sizable $F$-term.
The typical time scale of this phenomenon is given by the mass of the waterfall field. If this is
shorter than the modulus oscillation period at that time, the adiabaticity 
condition is violated, unless the modulus is coupled to the waterfall field
exactly in the same way as to the inflaton field. Thus, the adiabatic suppression mechanism
becomes inefficient and the modulus abundance is not exponentially suppressed.
%In order for the mechanism to work efficiently, the inflation model
%should be either 1) a single-field inflation model or 2) a multi-field inflation model with all the
%inflaton fields couple to the modulus in the same way.

In the following we study these issues in detail.

%%%%%%%%%%%%%%%%%%%%%%%%%%%%%%
\subsection{Polonyi/moduli production at the end of inflation}
\label{2-5-1}
%%%%%%%%%%%%%%%%%%%%%%%%%%%%%%

Let us first consider the simple model (\ref{Vmod}), with the initial
condition given at a finite Hubble parameter, $H_i = p/t_i$. 
The essential difference from  (\ref{supfac}) is that,
after inflation,  the potential minimum shifts from $z_*$ by a finite amount 
during the first period of oscillation $\sim (cH_i)^{-1}$. Note that the shift would be infinitely small if the
initial condition was given at $t=0$ when the  Hubble parameter is infinitely large (see \EQ{zmin}).
Once the modulus starts to oscillate about the time-dependent minimum,
the particle production is suppressed exponentially afterwards. So let us study the dynamics
during the first period of oscillation after the inflation.

Suppose that the minimum changes by $\delta z$
within $\delta t = \left(c H_i \right)^{-1}$ after inflation.
If the potential minimum changes proportionally to time in the first period of oscillation,
the initial amplitude is given by
\beq
\delta z (t_i) \;\sim\; \left|\dot{z}_{\rm min}(t_i)\right| \left(c H_i \right)^{-1},
\eeq
where $z_{\rm min}(t)$ denotes the time-dependent potential minimum,
and the dot means the derivative with respect to time.
The initial abundance of the modulus field created at the end of inflation is therefore given by
\beq
\frac{\rho_z^{(i)}}{s} \;\sim\;  \frac{T_{\rm R}}{8} \lrfp{z_{\rm min}(t_i) }{M_P}{2} \lrfp{\delta z(t_i)}{z_{\rm min}(t_i) }{2}.
\label{rhoz_i}
\eeq
In the model (\ref{Vmod}), we have
\bea
\label{zmin}
z_{\rm min}(t_i) &=&  \frac{c^2H_i^2}{m_z^2+c^2H_i^2} \,z_* \approx z_*, \\
\dot{z}_{\rm min}(t_i)&=&-\frac{3m_z^2 H_i}{m_z^2 + c^2 H_i^2}\, z_{\rm min}(t_i)
\approx -\frac{3m_z^2 }{ c^2 H_i} z_*,
\label{zdot}
\eeq
where we used $m_z \ll c H_i$ in the second equalities.
The inflaton-matter dominated phase is assumed in \EQS{rhoz_i} and \REF{zdot}.
(See discussion in \SEC{2-3}.)
The initial abundance for the model (\ref{Vmod}) is given by
\beq
\frac{\rho_z^{(i)}}{s} \;\sim\;  \frac{9}{8} T_{\rm R} \lrfp{z_*}{M_P}{2} \lrfp{m_z}{c H_i}{5}.   \label{rhoz_ii}
\eeq
Note that the condition  (\ref{condition}) is satisfied if $c \gg 1$, and therefore the particle production
afterward is exponentially suppressed. The final abundance is given by the sum of the estimate given in
Sec.~\ref{2-1} and \REF{rhoz_ii}.
Compared to $(\rho_z/s)_{\rm exp}$ given by \EQ{analytic1}, there is a suppression factor, $(m_z/c H_i)^5$
in \REF{rhoz_ii}.
Thus, the modulus abundance is power suppressed, if this additional contribution is dominant.

\begin{figure}[htbp]
\begin{center}
\includegraphics[scale=1.5]{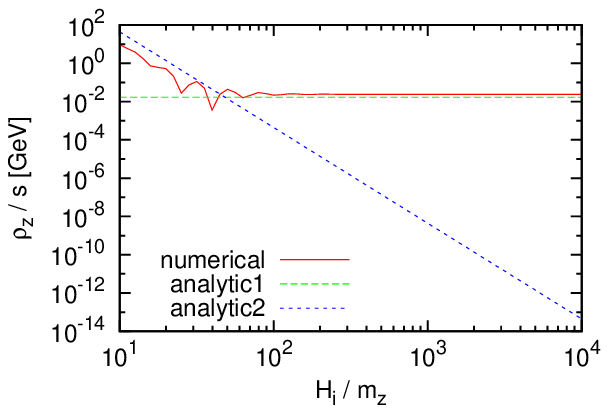}
\includegraphics[scale=1.5]{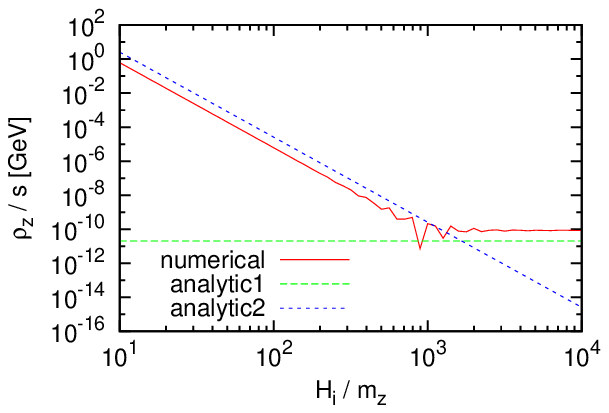}
\caption{
	The dependence of the modulus abundance on the initial condition $H_i$
	for $c=20$ (top) and $c=30$ (bottom).
	Numerical solutions are compared with analytic estimate of the 
	modulus abundance (\ref{supfac}) labeled as ``analytic1'', 
	and that of initial modulus abundance (\ref{rhoz_ii}) labeled as ``analytic2''.
	We have taken $\Gamma_\phi = 10^{-2.5}m_z$.
 }
\label{fig:Hi}
\end{center}
\end{figure}

We have checked that this analytic estimate agrees well with the numerical results.
In Fig.~\ref{fig:Hi}, the dependence of the modulus energy abundance on the initial condition $H_i$
for $c=20$ (top) and $c=30$ (bottom) are shown.
Numerical solutions are compared with analytic estimate of the modulus abundance 
based on (\ref{supfac}) labeled as ``analytic1'', 
and that of the initial modulus abundance based on (\ref{rhoz_ii}) labeled as ``analytic2''.
We have taken $\Gamma_\phi = 10^{-2.5}m_z$.
It is seen that analytic estimates well reproduce the numerical results
and that the initial modulus abundance (\ref{rhoz_ii}) becomes non-negligible as the ratio $H_i/m$ decreases.
This validates our consideration above.

%%%%%%%%%%%%%%%
%%%%%%%%%%%%%%%

In order for the above modulus abundance to be consistent with observations, 
the Hubble parameter during (precisely speaking, at the end of) inflation, 
$H_{\rm inf} \simeq H_{i}$, should satisfy
\beq
H_{\rm inf} \;\gsim 500 \,m_z \lrfp{c}{30}{-1}\lrfp{\Delta}{10^{-20}}{-\frac{1}{5}},
%517
\label{Hlow}
\eeq
where $\Delta$ denotes the required suppression factor for the modulus abundance
to be consistent with observation, normalized by $(\rho_z/s)_{\rm exp}$ in \EQ{analytic1}.
Although this inequality is satisfied for many inflation models, 
it certainly places a non-trivial lower bound on the inflation scale. The conditions (\ref{Hlow}) 
and (\ref{eq:upperTR}), together with $c \gg 1$,
are the necessary conditions for the adiabatic solution to work in the model (\ref{Vmod}).

In deriving (\ref{rhoz_ii}), we have implicitly assumed that the modulus does not ``see" the inflaton oscillations.
This is a valid assumption for low-scale inflation models such as new inflation models,
in which $m_\phi \gg H_{\rm inf}$ is satisfied 
where $m_\phi$ is the inflaton mass around the minimum.
In this class of models, the modulus minimum starts to move soon after the inflation ends, and the displacement during the first
modulus oscillation determines the initial abundance.
The particle production at the end of inflation can be understood because there is another time scale, i.e., the inflaton mass,
which is lighter than the modulus mass during inflation but becomes much heavier after the end of inflation.
As a result, the minimum to start moving in a time scale of $m_\phi$.  In other words,
the decoupling process of the inflaton violates the adiabaticity condition, leading to the particle production.

\begin{figure}[htbp]
\begin{center}
\includegraphics[scale=1.5]{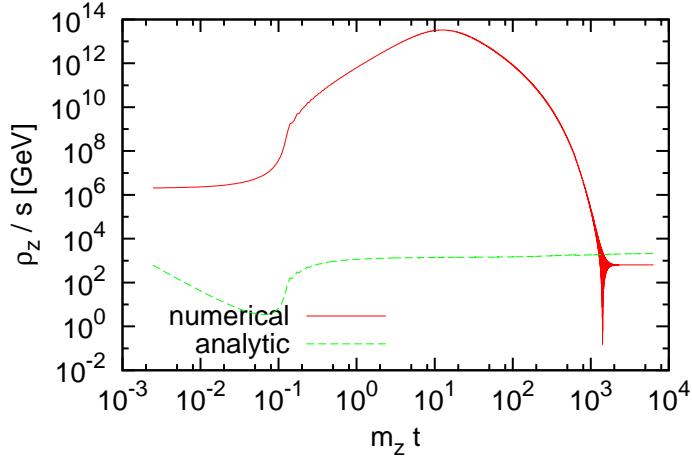}
\caption{
	The modulus abundance in the single-field chaotic inflation model.
	We have taken $c=20$, $m_\phi = 0.1M_P$, $m_z = 10^{-3}M_P$, $\Gamma_\phi=10^{-2.5}m_z$.
	Together shown is the analytic estimate based on (\ref{rhoz_i2}).
 }
\label{fig:single_chaotic}
\end{center}
\end{figure}

The situation is slightly different for the chaotic inflation model where $m_\phi \sim H_{\rm inf}$
and hence $m_\phi \ll c H_{\rm inf}$ for $c\gg 1$.
In this case the modulus remains heavier than the inflaton for a while after inflation, and
the adiabaticity is violated at $H \sim H_{\rm ad} \equiv m_\phi / c$.
The modulus abundance produced at $H_{\rm ad}$ is estimated in a similar manner
by replacing $H_i$ with $H_{\rm ad}$, and the result is
\beq
\frac{\rho_z^{(i)}}{s} \;\sim\;  \frac{9}{8} T_{\rm R} \lrfp{z_*}{M_P}{2} \lrfp{m_z}{m_\phi}{5}.   \label{rhoz_i2}
\eeq
We have confirmed that
this agrees with numerical calculation. See Fig.~\ref{fig:single_chaotic}.
Here the suppression factor is given by $9(m_z/m_\phi)^5$, with respect to  (\ref{analytic1}).
Recalling that the suppression factor of $\sim 10^{-20}$ is needed to solve the moduli problem,
and that the inflaton mass in the single-field chaotic inflation model is given by $m_\phi \simeq 2\times 10^{13}$\,GeV, 
the suppression factor $(m_z/m_\phi)^5$ is so small that the moduli produced in this way is cosmologically harmless.

%%%%%%%%%%%%%%%
%%%%%%%%%%%%%%%

%In SUSY, a single-field new inflation model~\cite{Izawa:1996dv} is a good example.
%In the multi-field inflation models, however, it is more difficult to solve the moduli problem 
%via the adiabatic suppression, as will be shown below.

%%%%%%%%%%%%%%%%%%%%%%%%%%%%%%%%%%%%
\subsection{Polonyi/moduli production in single-field inflation models}
\label{2-5-2}
%%%%%%%%%%%%%%%%%%%%%%%%%%%%%%%%%%%%

So far we have assumed that the position of the minimum does not jump
at the end of inflation. 
In supergravity based inflation models, however, this is not necessarily the case.
We show below that the minimum actually changes at the end of inflation in general single-field inflation models in supregravity,
including the model of Ref.~\cite{Izawa:1996dv}.

For illustration, let us consider the K\"ahler potential as
\begin{equation}
	K = |Z|^2 + |\phi|^2+ c^2 \frac{|Z-Z_*|^2 |\phi|^2}{M_P^2},
\end{equation}
where $c\gg 1$ and $\phi$ denotes the inflaton superfield.
We generically obtain the following form of the scalar potential from this K\"ahler potential,
\begin{equation}
	V = 3H_\phi^2 |z|^2 + c^2 H^2 |z-z_*|^2,   \label{Vsingle}
\end{equation}
where $3H_\phi^2 M_P^2 \equiv V_\phi$ with $V_\phi$ being the inflaton potential energy,
and $3H^2M_P^2 = |\dot \phi|^2 + V_\phi$.
The point is that the first term in (\ref{Vsingle}) depends only on the inflaton potential energy
and it rapidly oscillates after inflation.
The time averaged value of $H_\phi^2$ is given by $\langle H^2_\phi\rangle = H^2/2$
if the inflaton behaves as matter during the oscillation around its minimum.
Therefore, the position of the minimum during inflation is given by $z_{\rm min} = c^2 z_*/(c^2+3)$,
while the minimum after inflation is given by $z_{\rm min}' = c^2 z_*/(c^2+3/2)$.
Since the time scale of the oscillation of the $H_\phi^2$ term, which is determined by the inflaton mass,
is much greater than the modulus mass, the modulus does not feel the rapid oscillation of $H_\phi^2$ itself
but its time averaged value.
Thus the modulus oscillation is induced just after inflation with an amplitude of
\begin{equation}
	\delta z = |z_{\rm min}-z_{\rm min}'|=\frac{3z_*}{2c^2}.
\end{equation}
The resulting modulus abundance is estimated as
\beq
	\frac{\rho_z^{(i)}}{s} \;\sim\;  \frac{9}{32} T_{\rm R} \lrfp{z_*}{M_P}{2}
 	\left( \frac{m_z}{c^3 H_{\rm inf}} \right).    \label{rhoz_single}
\eeq
Compared with the standard estimation (\ref{analytic1}), the abundance is suppressed by the factor 
$\sim m_z/(c^3H_{\rm inf})$. 
This suppression is not sufficient, although the amount of tuning required for solving the moduli problem is
significantly relaxed.
For example, for $c=30$, $T_{\rm R}=10^6$\,GeV, $m_z = 1$\,TeV and $H_{\rm inf}=10^{12}$\,GeV,
the tuning of $|z_*| \lesssim 10^{-3}M_P$ is needed for solving the moduli problem.
The estimate (\ref{rhoz_single}) is similar to that in the multi-field inflation model discussed below.
We will show the validity of this estimate numerically in the next subsection.

%%%%%%%%%%%%%%%%%%%%%%%%%%%%%%%%%%%%
\subsection{Polonyi/moduli production in multi-field inflation models}
\label{2-5-3}
%%%%%%%%%%%%%%%%%%%%%%%%%%%%%%%%%%%%

Now let us discuss multi-field inflation models, where the position of the minimum could drastically change at the end
of inflation. The time scale of the change is usually determined by the mass of fields
acquiring the $F$-term after inflation, e.g., the mass of the waterfall fields in the hybrid inflation. 
Let us consider a class of inflation models with a superpotential
\begin{equation}
	W = X f(\phi).
\end{equation}
Most SUSY inflation models including hybrid inflation~\cite{Dvali:1994ms}, 
smooth-hybrid inflation~\cite{Lazarides:1995vr}, two-field new inflation~\cite{Asaka:1999jb}
chaotic inflation~\cite{Kawasaki:2000yn} and its variants~\cite{Takahashi:2010ky} fall into this category.\footnote{
	This form of the superpotential has been recently studied in detail in Ref.~\cite{Kallosh:2010ug}.
}
Here it is the $X$ whose $F$-term gives the inflaton potential energy during inflation.
In the two-field new and chaotic inflation models, $\phi$ is regarded as
the slowly-rolling inflaton field, while $X$ does not participate in the inflaton dynamics.
In the hybrid inflation model, $X$ plays the role of the slowly-rolling inflaton, while $\phi$ is the waterfall field.

In general, $X$ and $\phi$ can have different couplings to the modulus field,
\begin{equation}
	K = \tilde{c}_X^2 \frac{|X|^2 |Z-Z_X|^2}{M_P^2} +  \tilde{c}_\phi^2 \frac{|\phi|^2 |Z-Z_\phi|^2}{M_P^2},   
	\label{Kahler_multi}
\end{equation}
in the K\"ahler potential. This yields the scalar potential for the modulus as\,\footnote{
	There are other terms for the modulus potential, 
	but (\ref{Vsugra_multi}) is dominant if $c_X^2, c_\phi^2 \gg 1$ and if the other terms
	are suppressed by the Planck scale with coefficients of order unity.
}
\begin{equation}
	V =  \frac{1}{M_P^2}\left[  \tilde{c}_X^2 \left( |\dot X|^2 + |F_X|^2\right) \left|Z-Z_X\right|^2 
		+ \tilde{c}_\phi^2\left( |\dot\phi|^2 + |F_\phi|^2\right)  \left|Z-Z_\phi\right|^2 \right],
	\label{Vsugra_multi}
\end{equation}
where $X$ and $Z$ represent the lowest component of the corresponding superfields.
During inflation it is $F_X$ that contributes to the Hubble parameter, while $F_\phi$ and
the kinetic energy of $X$ and $\phi$ are subdominant or negligibly small.  In the two-field new and chaotic inflation models, 
$X$ sits at the origin and does not participate in the dynamics. 
After the inflation ends, both $|\dot{\phi}|^2$ and  $|F_X|^2$ oscillate with the same amplitude and opposite phase,
while $|\dot{X}|^2$ and $|F_\phi|^2$ remain negligible.
On the other hand, in the hybrid inflation model, all of them oscillates with a time scale of the inflaton mass
in a complicated manner. 

In order to take into such behavior in the multi-field inflation models,
let us parametrize the modulus  potential  as
\begin{equation}
	V = \frac{1}{2}m_z^2 z^2 + \frac{1}{2}c_X^2 H_1^2 (z-z_X)^2 + \frac{1}{2}c_\phi^2 H_2^2 (z-z_\phi)^2,
	\label{Vz_multi}
\end{equation}
where the Hubble parameter is divided into two parts as $H^2 = H_1^2 + H_2^2$.
We assume that $H_1^2 \simeq H^2$ and $H_2^2 \simeq 0$ during inflation and that $H_1^2$ and
$H_2^2$ oscillate with an opposite phase after inflation. Note that their sum is equal to the Hubble
parameter and so it does not oscillate. The time-averaged values of $H_1^2$ and $H_2^2$
are given by
\beq
\la H_1^2 \ra = \la H_2^2 \ra = \frac{H^2}{2}.
\label{ta}
\eeq
During inflation, where the kinetic energy of the inflaton is negligible, the modulus sits at $z=z_X$.
After inflation ends, the temporal minimum oscillates around between $z_X$ and $z_\phi$,
and it is given by
\begin{equation}
	z_{\rm min} = \frac{c_X^2 H_1^2 z_X + c_\phi^2 H_2^2 z_\phi }
	{c_X^2 H_1^2 + c_\phi^2 H_2^2}.
	\label{zmin2}
\end{equation}
Its time derivative is 
\begin{equation}
	\dot z_{\rm min} = \frac{2c_X^2 c_\phi^2 H_1H_2(H_1\dot H_2+H_2\dot H_1) }
	{(c_X^2 H_1^2 + c_\phi^2 H_2^2)^2} (z_X-z_\phi),
\end{equation}
where we have approximated $m_z^2 \ll c_X^2 H_1^2 + c_\phi^2 H_2^2$.
Hereafter we consider the case of $c_X = {\cal O}(10)$ and $c_X \gg c_\phi$ since otherwise
it is obvious that the coherent oscillations of the modulus with an amplitude of $\sim z_X$ or $z_\phi$
is induced at the end of inflation, and the cosmological moduli problem is not solved
if $z_X \sim z_\phi \sim M_P$.

Let us first consider the case of $c_X H_{\rm inf} \ll m_\phi$, which is satisfied 
in the hybrid and new inflation models.
 In this case the modulus mass is much smaller than the typical frequency of the inflaton oscillations,
and so it  only feels the time-averaged potential. At the end of inflation, the potential minimum changes instantly (in a
time scale of $m_\phi^{-1}$) from $z_X$ to $\bar z_{\rm min}$, where the time-averaged potential minimum,
$\bar z_{\rm min}$ is given by
\beq
\bar z_{\rm min} \;\simeq\; (c_X^2 z_X+c_\phi^2 z_\phi)/(c_X^2+c_\phi^2).
\eeq
Here we have used \EQS{ta} and \REF{zmin2}.
Thus the initial amplitude of the modulus oscillation is
\begin{equation}
	\delta z(t_i) \sim z_X - \bar z_{\rm min} \sim \frac{c_\phi^2}{c_X^2}(z_X-z_\phi).
\end{equation}
The modulus abundance produced at the end of inflation in this case is estimated as
\beq
	\frac{\rho_z^{(i)}}{s} \;\sim\;  \frac{1}{8} T_{\rm R} \lrfp{z_X-z_\phi}{M_P}{2}
 	\left( \frac{c_\phi^4 m_z}{c_X^3 H_{\rm inf}} \right).    \label{rhoz_i3}
\eeq
Although this may be still too large to avoid the cosmological Polonyi/moduli problem,
the abundance is suppressed by a huge factor with respect to the original case without the adiabatic suppression mechanism.
For example, for $c_X=30$, $c_\phi=1$, $T_{\rm R}=10^6$\,GeV, $m_z = 1$\,TeV and $H_{\rm inf}=10^{12}$\,GeV,
the tuning of $|z_X-z_\phi | \sim 10^{-3}M_P$ is sufficient for solving the moduli problem.

\begin{figure}[htbp]
\begin{center}
\includegraphics[scale=1.5]{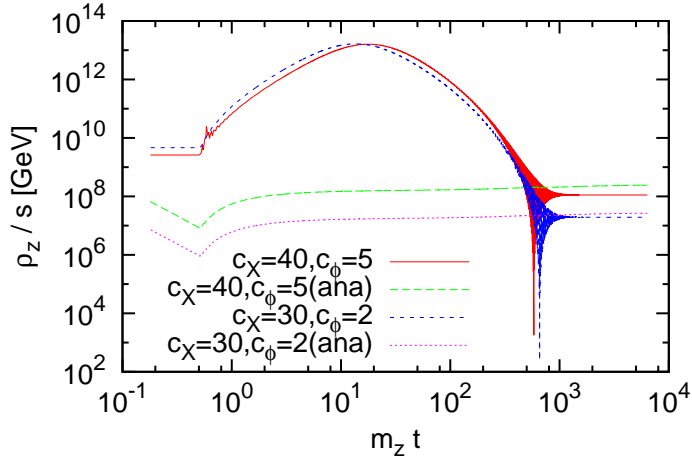}
\caption{
	The modulus abundance in the hybrid inflation model.
	We have taken $\kappa=1$, $\mu=0.1M_P$, $m_z = 10^{-3}M_P$, $\Gamma_\phi=10^{-2.5}m_z$
	and $z_X=M_P/c_X$ and $z_\phi = z_X/2$.
	Results for  $(c_X,c_\phi)=(30,2)$ and $(40,5)$ are shown 
	together with the analytic estimate based on (\ref{rhoz_i3}).
 }
\label{fig:multi_hybrid}
\end{center}
\end{figure}

In order to check this estimate, we have evaluated the modulus abundance in the SUSY hybrid inflation model
with the superpotential~\cite{Dvali:1994ms},
\begin{equation}
	W = \kappa X (\phi \bar \phi- \mu^2).
\end{equation}
For simplicity we assume the following form in the potential (\ref{Vz_multi}),
\begin{eqnarray}
	H_1^2 &=& \frac{\kappa^2 \left| \mu^2 - \phi \bar\phi \right|^2 + |\dot X|^2}{3 M_P^2}, \\
	H_2^2 &=& \frac{\kappa^2 |X|^2( |\phi|^2 + |\bar \phi|^2) + |\dot \phi|^2 + |\dot {\bar\phi}|^2}{3 M_P^2} .
\end{eqnarray}
The modulus abundance in this model is shown in Fig.~\ref{fig:multi_hybrid}.
We have taken $\kappa=1$, $\mu=0.1M_P$, $m_z = 10^{-3}M_P$, $\Gamma_\phi=10^{-2.5}m_z$
and $z_X=M_P/c_X$ and $z_\phi = z_X/2$.
Results  for  $(c_X,c_\phi)=(30,2)$ and $(40,5)$ are shown
together with the analytic estimate based on (\ref{rhoz_i3}).
We can see that the analytic estimate on the final modulus abundances agree well with the numerical results.

%%%%%%%%%%%%%%%
%%%%%%%%%%%%%%%

\begin{figure}[htbp]
\begin{center}
\includegraphics[scale=1.5]{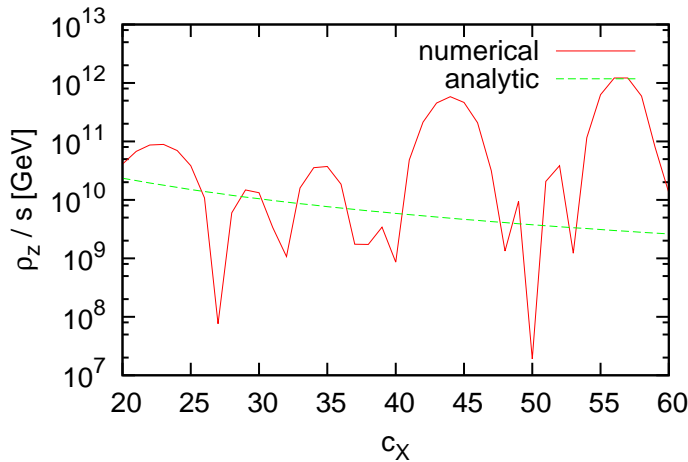}
\caption{
	The modulus abundance as a function of $c_X$ in the multi chaotic inflation model.
	We have taken $m_\phi=0.1M_P$, $m_z = 10^{-3}M_P$, $\Gamma_\phi=10^{-2.5}m_z$, $c_\phi=2$
	and $z_X=M_P/c_X$ and $z_\phi = z_X/2$.
	Analytic estimate based on (\ref{rhoz_i4}) is also shown.
 }
\label{fig:multi_chaotic}
\end{center}
\end{figure}

Next, we consider the other case : $H_{\rm inf} \sim m_\phi$ 
as in the chaotic inflation model in supergravity~\cite{Kawasaki:2000yn}.
The superpotential is given by
\beq
W \;=\; m_\phi X \phi,
\eeq
and the inflaton $\phi$ starts to oscillate from about the Planck scale.
The $X$ is stabilized at the origin during and after inflation.
We therefore take
\bea
H_1^2 &=& \frac{m_\phi^2 |\phi|^2}{3 M_P^2},\\
H_2^2  &=& \frac{|\dot \phi|^2}{3 M_P^2}.
\eea
In this case,  the modulus oscillates in a potential which changes with a
frequency $m_\phi^{-1}$. Since the modulus mass
and the inflaton mass become comparable for a certain time, 
it is likely that resonant particle production occurs, making it difficult to estimate the modulus
 abundance analytically. Let us therefore give a very rough estimate on the abundance
 of the moduli in the form of coherent oscillations.  We assume $c_X = {\cal O}(10)$
 and $c_\phi = {\cal O}(1)$. 
Since both $H_1$ and $H_2$ oscillates at a frequency of $m_\phi$ with an opposite phase,
the modulus mass becomes of order the Hubble parameter when $H_1$ vanishes. During the time
of $H^{-1} \sim m_\phi$, the modulus moves from  $z_X$ to $\sim z_\phi$. As the modulus mass becomes
greater than the Hubble parameter, it comes back to $z_X$ again and starts oscillating with an initial
amplitude given by $\sim |z_\phi - z_X|$.
Therefore, the Polonyi abundance in the form of coherent oscillations is estimated as
\beq
	\frac{\rho_z^{(i)}}{s} \;\sim\;  \frac{1}{8} T_{\rm R} \lrfp{z_X-z_\phi}{M_P}{2}
 	\left( \frac{m_z}{m_\phi} \right)     \label{rhoz_i4}
\eeq
for  $c_X = {\cal O}(10)$  and $c_\phi = {\cal O}(1)$. 
Numerical results are shown in Fig.~\ref{fig:multi_chaotic}.
Here the modulus abundance is evaluated as a function of $c_X$ in the multi chaotic inflation model.
We have taken $m_\phi=0.1M_P$, $m_z = 10^{-3}M_P$, $\Gamma_\phi=10^{-2.5}m_z$, $c_\phi=2$
and $z_X=M_P/c_X$ and $z_\phi = z_X/2$.
One can see that there is a resonant feature and the analytic estimate roughly agrees with the numerical
result.  Note however that we have focused on the homogeneous mode of the modulus, and
the result might be significantly affected by taking account of the resonant particle production, or
preheating. Assuming that the resonant particle production, if any, is just added to the above estimate, 
the modulus abundance is huge and another solution to the cosmological Polonyi/moduli problem 
is needed.\footnote{
If $c_\phi$ is much smaller than unity for some reason, the modulus abundance is likely suppressed. 
}

%%%%%%%%%%%%%%%
%%%%%%%%%%%%%%%

After all, in the multi-field inflation models,
the adiabatic suppression is inefficient once the modulus production at the end of inflation is taken into account.
We need to tune the couplings of the multi-inflaton fields to the Polonyi/moduli 
so that $z_X\simeq z_\phi$ in order to solve the cosmological Polonyi/moduli problem,
although the required amount of tuning is relaxed from $10^{-10}$ to $10^{-3}$
compared with the case of $c=\mathcal O(1)$.

%%%%%%%%%%%%%%%%%%%%%%%%%%%%%%%%%%%%
\subsection{Required fine-tuning to solve the Polonyi/moduli problem}
\label{sec:tuning}
%%%%%%%%%%%%%%%%%%%%%%%%%%%%%%%%%%%%

\begin{figure}[htbp]
\begin{center}
\includegraphics[scale=1.0]{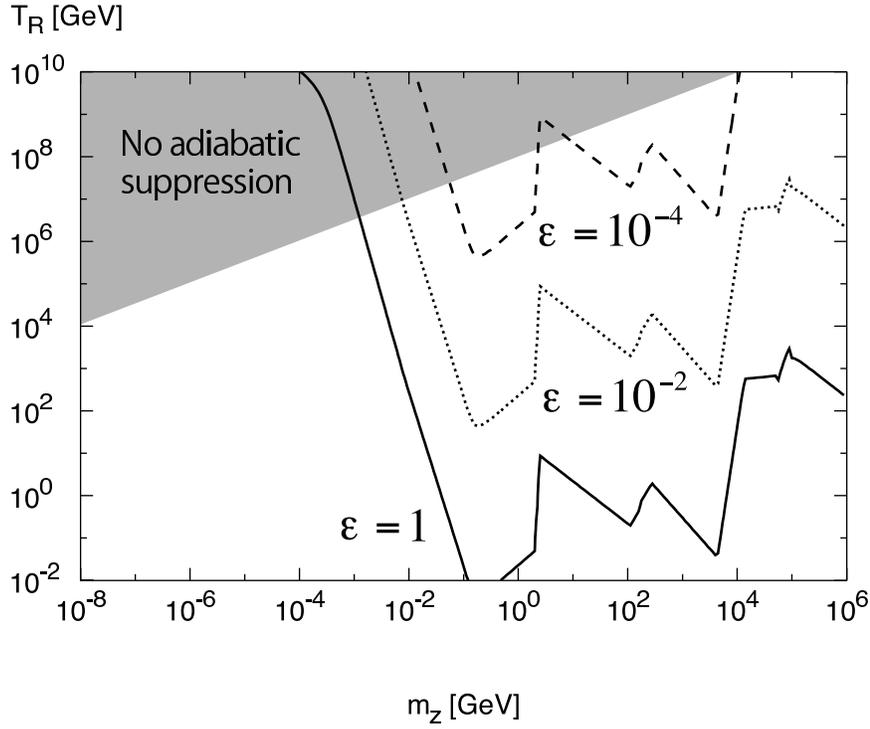}
\caption{
	Contours of the required amount of the fine-tuning $\epsilon$ for the adiabatic
	suppression to solve the moduli problem,
	on the plane of $m_z$ and $T_{\rm R}$.
	Here we have taken $H_{\rm inf}=10^{10}$\,GeV and $c_X=30$, $c_\phi=1$.
	The required amount of tuning becomes less (more) stringent 
	by one order of magnitude for $H_{\rm inf}=10^{12}$ $(10^{8})$\,GeV.
	That is to say, $\epsilon$ scales as $(H_{\rm inf}/10^{10}{\rm GeV})^{1/2}$.
	In the shaded region at the upper left corner, the adiabatic suppression does not work.
	Notice that upper bounds on the reheating temperature coming from 
	Polonyi/moduli and gravitino thermal production, which will be discussed in the following sections,
	 are not taken into account in this figure. 
 }
\label{fig:tuning}
\end{center}
\end{figure}

In order to see the required amount of tuning more quantitatively, 
we define the tuning factor $\epsilon$ as $|z_X-z_\phi|\equiv \epsilon M_P$
and examine how small $\epsilon$ must be in order to solve the cosmological moduli problem.
To be concrete, we rely on the estimate (\ref{rhoz_i3}).
This estimate applies almost identically to the case of single-field inflation models if we set $c_\phi=1$,
as shown in Eq.~(\ref{rhoz_single}).
Fig.~\ref{fig:tuning} shows the contours of the required amount of the fine-tuning $\epsilon$ 
for the adiabatic suppression to solve the moduli problem, on the plane of $m_z$ and $T_{\rm R}$.
Here we have taken $H_{\rm inf}=10^{10}$\,GeV and $c_X=30$, $c_\phi=1$.
The required amount of tuning becomes less (more) stringent 
by one order of magnitude for $H_{\rm inf}=10^{12}$ $(10^{8})$\,GeV.
That is to say, $\epsilon$ scales as $(H_{\rm inf}/10^{10}{\rm GeV})^{1/2}$.
In the shaded region at the upper left corner, the adiabatic suppression does not work.
We have used constraints from BBN, diffuse X($\gamma$)-ray and overclosure as done in 
Ref.~\cite{Asaka:1999} with updated BBN constraints.
For $m_z > 100$\,TeV, we have assumed anomaly mediation relation between the 
gravitino and the Wino-LSP mass and derived constraints from LSP overproduction by the modulus decay.
We assumed that the modulus has couplings with the SM particles as Eq.~(\ref{Zgg}) with $c_3=1$.
We have also assumed that the modulus mass is comparable to the gravitino mass so that the
gravitino production from the modulus decay is inefficient.
Notice that upper bounds on the reheating temperature coming from 
Polonyi/moduli and gravitino thermal production, which will be discussed in the following sections,
are not taken into account in this figure. 
From this figure, it is seen that for wide range of modulus mass and reheating temperature,
the required amount of tuning lies in the range of $10^{-4} \lesssim \epsilon \lesssim 1$.

%%%%%%%%%%%%%%%
%%%%%%%%%%%%%%%

%%%%%%%%%%%%%%%%%%%%%%%%%%%%%%
\subsection{Other form of the modulus potential}
%%%%%%%%%%%%%%%%%%%%%%%%%%%%%%

Before closing this section, we comment on whether the adiabatic suppression works
in other types of the modulus potential.
First we consider the case of a negative Hubble mass squared. The scalar potential is 
\begin{equation}
	V(z) = \frac{1}{2}(m_z^2 - c^2 H^2)z^2 + \frac{\lambda}{n M_P^{n-4}} z^n,    \label{pot2}
\end{equation}
where  $n (\geq 4)$ is an even integer and $\lambda$ is a positive constant.
The flat directions in SSM have the potential of this type~\cite{Affleck:1984fy}.
The true minimum is obviously $z=0$.
The temporal minimum is given by
\begin{equation}
	z_{\rm min}(H) = \left \{ \begin{array}{ll}
	\left [ M_P^{n-4} (c^2 H^2 - m_z^2)/\lambda \right]^{1/(n-2)} & ~~~{\rm for}~~~cH > m_z \\
	0 & ~~~{\rm for}~~~cH < m_z.
	\end{array}
	\right. .
\end{equation}
Hence we obtain
\begin{equation}
	f(H) = \frac{3}{n-2}\frac{c^2 H^3}{c^2 H^2 - m_z^2}~~~{\rm for}~~~cH > m_z,
\end{equation}
in the matter-dominated phase.
Therefore,  $f(H)$ diverges at $H=m_z/c$
and the adiabaticity is always violated independently of the size of $c$ and $\lambda$.
Thus the adiabatic suppression does not occur for the scalar field of the potential of (\ref{pot2}).
The $z$ field begins to oscillate at $H\sim m_z/c$ around the minimum $z=0$
with amplitude of $\sim (M_P^{n-4}m_z^2/\lambda)^{1/(n-2)}$.
We have numerically confirmed that the oscillation amplitude is not suppressed
regardless of the value of $c$.

\begin{figure}[htbp]
\begin{center}
\includegraphics[scale=1.5]{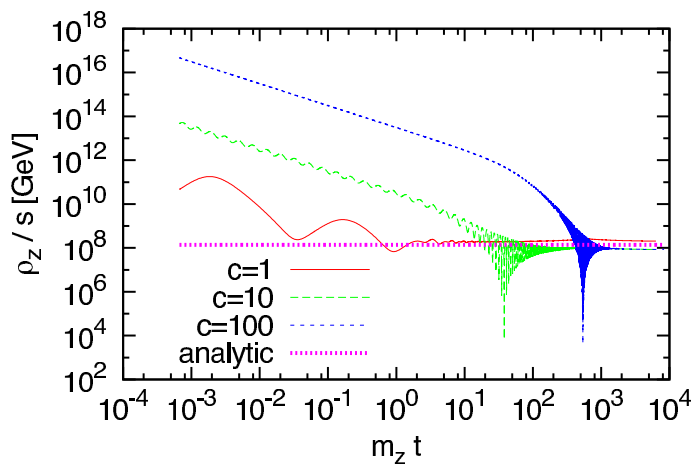}
\caption{
	Time evolution of the modulus abundance for $c=1, 10, 100$
	for the model (\ref{pot3}). We have taken $n=6$ and $\lambda = M_P^2/m_z^2=10^6$,
	$\Gamma_\phi=10^{-2.5}m_z$ and $H_i=10^3 m_z$.
 }
\label{fig:model3}
\end{center}
\end{figure}

The situation slightly changes if $z$ has the negative mass term around the origin :
\begin{equation}
	V(z) = \frac{1}{2}(-m_z^2 - c^2 H^2)z^2 + \frac{\lambda}{n M_P^{n-4}} z^n.    \label{pot3}
\end{equation}
The temporal minimum is given by
\begin{equation}
	z_{\rm min}(H) =
	\left[ M_P^{n-4} (c^2 H^2 + m_z^2)/\lambda \right]^{1/(n-2)} .
\end{equation}
It smoothly connects to the true minimum 
$z_{\rm min} = \left( M_P^{n-4} m_z^2/\lambda \right)^{1/(n-2)} $.
In this case we obtain
\begin{equation}
	f(H) = \frac{3}{n-2}\frac{c^2 H^3}{c^2 H^2 + m_z^2},
\end{equation}
in the matter-dominated phase, hence the time scale of the potential change is roughly given by $H$.
This is always smaller than the mass scale around the temporal minimum for $c\gg 1$
and hence the adiabaticity condition (\ref{condition}) is met in the oscillation regime. 
As is already explained, however, $z$ cannot track the temporal minimum at the very beginning of its motion.
Thus the coherent oscillation is necessarily induced as explained before.
The induced modulus amplitude is estimated as
\begin{equation}
	\delta z(t_i) \sim |\dot z_{\rm min} (H_i)| (cH_i)^{-1}  \sim z_{\rm min}(H_i)/c,
\end{equation}
and the resulting modulus abundance is
\begin{equation}
	\frac{\rho_z^{(i)}}{s} \sim \frac{1}{8}T_{\rm R} \left( \frac{z_{\rm min}(H_i)}{M_P} \right)^2 
	\left( \frac{m_z}{cH_i} \right).
\end{equation}
This is the minimum modulus abundance in this model.
By tuning the initial velocity $\dot z$, we can suppress the abundance further,
but such a tuning is not likely to occur in a realistic setup starting from inflation.
Numerical calculations and analytic estimate are shown in Fig.~\ref{fig:model3}.
We have taken $n=6$ and $\lambda = M_P^2/m_z^2=10^6$
and $H_i=10^3 m_z$. Analytic estimate does not depend on $c$ for $n=6$.
Our estimate fits very well with the numerical results.

%%%%%%%%%%%%%%%
%%%%%%%%%%%%%%%

\begin{figure}[htbp]
\begin{center}
\includegraphics[scale=1.5]{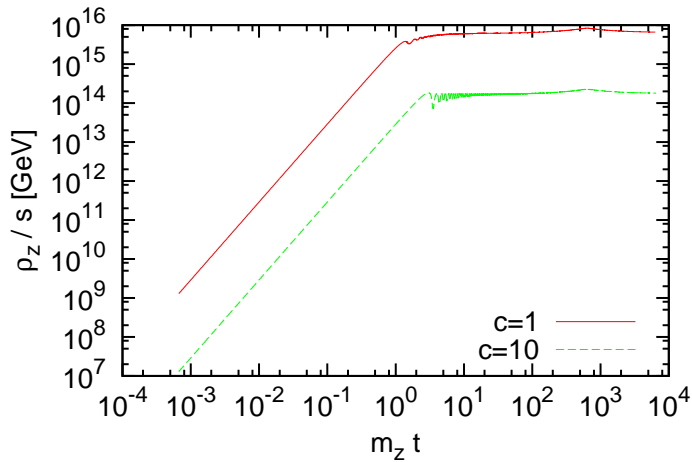}
\caption{
	Time evolution of the modulus abundance for $c=1, 10$
	for the model (\ref{pot4}). 
	We have taken $z_*=M_P/c$, $M=z_*/10$ and $V_0 = m^2 z_*^2$ with $m=10^{-3}M_P$.
 }
\label{fig:model4}
\end{center}
\end{figure}

The last example is a very flat potential such as~\cite{Lukas:1996wx}
\beq
	V = V_0 \log \left( 1+ \frac{z^2}{M^2} \right) + \frac{1}{2}c^2 H^2 (z-z_*)^2.   \label{pot4}
\eeq
Time evolution of the modulus abundance for $c=1, 10$
for the model (\ref{pot4}) is shown in Fig.~\ref{fig:model4}.
We have taken $z_*=M_P/c$, $M=z_*/10$ and $V_0 = m^2 z_*^2$ with $m=10^{-3}M_P$.
Adiabatic suppression does not work in this case, simply because the mass around the temporal minimum changes the sign
at some $z < z_*$ and the adiabaticity is necessarily violated.
If, instead, we have chosen $M \gg z_*$, the effective mass is always positive for $z<z_*$
and the model is effectively described by (\ref{Vmod}), hence the adiabatic suppression occurs for $c \gg 1$.

%%%%%%%%%%%%%%%
%%%%%%%%%%%%%%%

%%%%%%%%%%%%%%%%%%%%%%%%%%%%%%%%%%%%
\subsection{Summary}
\label{sec:3sum}
%%%%%%%%%%%%%%%%%%%%%%%%%%%%%%%%%%%%

To summarize, it depends on the behavior of the time-dependent potential minimum
whether the adiabatic suppression takes place. In particular,  if the curvature of the potential
about the minimum vanishes temporarily, or if the position of the minimum changes rapidly
compared to the curvature, the adiabatic solution does not work. Therefore it should be kept
in mind that the adiabatic solution does not apply to any moduli with an arbitrary potential.
Nevertheless, it is remarkable that the adiabatic suppression mechanism works for the simplest 
example like (\ref{Vmod}), which includes the Polonyi field.

It should be noted, however, that a considerable amount of modulus oscillation is generically induced at the end of inflation,
as already shown. 
Some amount of tuning is needed in order to solve the cosmological moduli problem for relatively high reheating temperature, 
although the required amount of tuning
is significantly reduced compared with the ordinary case without the adiabatic suppression.

In the following sections we always assume that the modulus/Polonyi coherent oscillation is somehow suppressed
by the adiabatic suppression and some amount of tuning if needed and discuss 
how the reheating temperature is constrained even in such a case.

%%%%%%%%%%%%%%%%%%%%%%%%%%%%%%%%%%%%
\section{The Polonyi problem in gravity mediation}
\label{sec:3}
%%%%%%%%%%%%%%%%%%%%%%%%%%%%%%%%%%%%

Let us consider the Polonyi problem in gravity mediation, where
SUSY is broken by the F-term of the  Polonyi field $Z$.
We consider the following K\"ahler potential,
\begin{equation}
	K = |\phi|^2+|Z|^2- c_1^2 \frac{ |\phi|^2 |Z-Z_*|^2 }{M_P^2} 
	-\frac{c_2^2}{4}\frac{|Z|^4}{M_P^2},   \label{PPZZ}
\end{equation}
where $Z_*$ represents the potential minimum during inflation.
Here we have omitted interactions like $K \supset \kappa M_P Z + \kappa' |\phi|^2 ZZ /M_P^2+ \cdots$.
The linear term is  necessary to set the origin to be the low-energy minimum, which also
shifts the minimum during inflation from $Z_*$ by some factor, but not an order of magnitude. 
So we drop the linear term since it does not change the modulus dynamics significantly in our context. 
The latter coupling induces the inflaton decay into a pair of the gravitinos~\cite{Kawasaki:2006gs}, 
which may impose severe constraints on inflation models if $\kappa' \gg 1$.

%Notice that in general there should be an inflaton self-coupling term in the K\"ahler potential
%which might spoil the inflationary dynamics.
%But in the multi-field inflation model, the field $\Phi$ that gives a large Hubble mass to the Polonyi
%during the inflaton oscillation does not always coincide with the true inflaton field 
%that slowly rolls down the potential.
%In this case, enhanced couplings of $\Phi$ do not necessarily mean 
%the introduction of the $\eta$-problem~\cite{Takahashi:2010uw}.

The coupling with the inflaton in the K\"ahler potential (\ref{PPZZ}) induces a Hubble mass correction to 
the effective potential of the Polonyi field as
\beq
V(z) \; \supset \; c^2 H^2 |z-z_*|^2
\eeq
where $z$ denotes a scalar component of $Z$, with
\begin{equation}
	c^2 = 3(c_1^2+1),
\end{equation}
which is equivalent to Eq.~(\ref{Vmod}). 
The Hubble-induced mass term is present during inflation and the inflaton oscillation era, but it
 disappears after the reheating. 
The Polonyi mass around the origin in the low energy is~\footnote{
Precisely speaking, this is the mass of the real component of $Z$ 
in the presence of the linear term with $\kappa = \sqrt{3} - 1$. The imaginary part has
a slightly smaller but comparable mass. 
} 
\begin{equation}
	m_z^2 = (3c_2^2 + 2\sqrt{3})m_{3/2}^2,
\end{equation}

In order to suppress the coherent oscillations of the Polonyi field,
 we assume $c ={\cal O}(10)$ and $T_{\rm R} \lsim 0.05 \sqrt{m_z M_P}$. These two conditions
must be satisfied for the adiabatic suppression to work, independently of the details of the inflaton dynamics.
%In the following we do not consider the Polonyi production at the end of inflation because it significantly depends on inflation models.
As shown in the previous section, 
the Polonyi coherent oscillation is generically induced at the end of inflation. 
Hereafter we simply assume that this contribution is suppressed 
at the price of fine-tuning of at most $10^{-4}$ level (see Fig.~\ref{fig:tuning}).

The Polonyi couplings to the SSM fields are parametrized as
\begin{equation}
\int d^2 \theta\, \frac{1}{4}\left(1+c_3\frac{Z}{M_P} \right)W_a W^a + {\rm h.c.} ,  \label{Zgg}
\end{equation}
\begin{equation}
	\int d^4 \theta\, \left\{-c_4^2\frac{ |Z|^2 | f |^2 }{M_P^2}  + \left( \frac{c_5}{M_P}Z |f|^2 + {\rm h.c.} \right) \right\}.  \label{zzff}
\end{equation}
where $W_a$ is a field strength of the gauge supermultiplet, $f$ collectively denotes the chiral matter superfield,
and $c_3$, $c_4$ and $c_5$ are numerical coefficients. In the following
we take all the coupling constants $c_1, \cdots ,c_5$ to be real, for simplicity. 
These terms generate the soft SUSY breaking masses as
\bea
	m_{\tilde g} &=& \frac{{\sqrt 3} c_3}{2}m_{3/2}, \\
	m_{\tilde f}^2 &=& (c_4^2+c_5^2+1) m_{3/2}^2,
\eea
where $m_{\tilde g}$ and $m_{\tilde f}$ denote the gaugino and sfermion masses, respectively.
The above couplings  enable the Polonyi field to decay into the SSM particles, and more important, 
the Polonyi is necessarily produced from particle scattering in thermal plasma through the couplings with
$c_3$ and $c_5$. (Note that the origin of $Z$ is set to be the low-energy potential minimum.) 
Therefore, even if the coherent oscillations of the Polonyi field can be negligibly small
by the adiabatic suppression mechanism, 
its abundance may be still non-negligible if the reheating temperature is high. 
%Therefore it is highly non-trivial whether the adiabatic solution makes the thermal leptogenesis
%viable or not.  

In the rest of this section, we consider cases with several different values of $c_i$.
First we consider the case that only $c_1$ is enhanced while all the other couplings are 
of order unity. This is the minimal set-up to solve the Polonyi problem
using the adiabatic solution. Next we extend the minimal set-up to allow enhancement of other couplings.
Such extension  may be indeed reasonable;
if the enhanced coupling of $c_1$ is due to an exchange of fields of mass below the Planck scale
or due to some strong dynamics at the Planck scale,
we naively expect a coupling like $c_2$ 
%as well as $|\Phi|^4$ 
is similarly enhanced.
Furthermore, if the SSM particles are involved with the
strong coupling at the Planck scale, or if the fundamental cut-off scale of theory is one order of magnitude
smaller than the Planck scale, we expect that all the modulus couplings to the SSM particles, $c_3$, $c_4$ and $c_5$ as well
are  enhanced. As we will see below, 
the constraints on the reheating temperature depends on the size of these couplings constants.

%%%%%%%%%%%%%%%% table %%%%%%%%%%%%%%%%%%%%%%
\begin{table}[t]
  \begin{center}
    \begin{tabular}{ | c || c | l | }
      \hline 
          parameter & definition & effects  \\ \hline \hline
          $c_1$  & Eq.~(\ref{PPZZ}) & Hubble mass for the Polonyi \\ \hline
          $c_2$  & Eq.~(\ref{PPZZ}) & Polonyi mass  \\  \hline
          $c_3$  & Eq.~(\ref{Zgg})    & Gaugino mass, Polonyi abundance  \\ \hline
          $c_4$  & Eq.~(\ref{zzff})    & Sfermion mass \\ \hline
          $c_5$  & Eq.~(\ref{zzff})    & Sfermion mass, Polonyi abundance \\ \hline
    \end{tabular}
    \caption{ 
    	Coefficients of non-renormalizable operators and their effects on masses and the
	Polonyi abundance from thermal scattering.
     }
  \end{center}
  \label{table:c}
\end{table}
%%%%%%%%%%%%%%%%%%%%%%%%%%%%%%%%%%%%%%%%%%%%%% 

%%%%%%%%%%%%%%%%%%%%%%%%%%%%%%%%%%%%
\subsection{The minimal set-up}
%%%%%%%%%%%%%%%%%%%%%%%%%%%%%%%%%%%%

First we consider a case that the Polonyi-inflaton coupling ($c_1$) is enhanced  to solve the Polonyi problem, 
while all the other couplings among the Polonyi and SSM particles are assumed to be Planck-suppressed
with coefficients of order unity. 
%We will briefly comment on the case of $c_3 \ll 1$ and $c_5 \ll 1$ later.

The interaction (\ref{Zgg}) induces the Polonyi decay into the SM gauge bosons,
which is subject to stringent constraints from BBN~\cite{Kawasaki:2004qu,Jedamzik:2006xz,Cyburt:2010vz}.
The partial decay rate is estimated  as
\begin{equation}
	\Gamma_{Z\to gg} \simeq  c_3^2\frac{3m_z^3}{32\pi M_P^2} 
	\simeq (1.3\times 10^5{\rm sec})^{-1} \,c_3^{2} \lrfp{m_z}{1{\rm\,TeV}}{3},
	\label{decay_Zgg}
\end{equation}
where we have considered the decay into all the SM gauge bosons.
The partial decay rates of the Polonyi into SSM gauginos is given by, if kinematically allowed,
\begin{equation}
	\Gamma_{Z\to \tilde g\tilde g} \simeq  c_3^2\frac{3m_z}{32\pi M_P^2}(m_{3/2}^2 + m_{\tilde g}^2),
	\label{decay_Zgaugino}
\end{equation}
where we have approximated all the gaugino masses are same.
The partial decay rates of the Polonyi into sfermions is also close to the above rate. 
Note that, if the Polonyi mass is much heavier than the gravitino mass (i.e. $c_2 \gg 1$),
the decay into the gauge bosons is the dominant decay mode~\cite{Endo:2006ix}.
%In the following analysis we assume that the total decay rate of the Polonyi field is approximately
%given by  (\ref{Zgg}).

The Polonyi is necessarily produced by thermal scattering like the gravitino,
even if the coherent oscillations are suppressed by the adiabatic suppression mechanism. 
The gravitino abundance is~\cite{Moroi:1993mb,Bolz:2000fu}
\begin{equation}
	Y_{3/2} \equiv \frac{n_{3/2}}{s} \sim 2\times 10^{-12}\left( 1+\frac{m_{\tilde g}^2}{3m_{3/2}^2} \right)
	\left( \frac{T_{\rm R}}{10^{10}{\rm GeV}} \right),
	\label{gravTP}
\end{equation}
where $n_{3/2}$ is the number density of the gravitino, and $s$ the entropy density. The Polonyi abundance is expected to be
the same order of that of the transverse component of the gravitinos. So we use the following estimate
\begin{equation}
	Y_{Z} \equiv \frac{n_Z}{s} \sim 10^{-12} (c_3^2 + c_5^2) \left( \frac{T_{\rm R}}{10^{10}{\rm GeV}} \right),
	\label{modTP}
\end{equation}
where $n_Z$ is the number density of the moduli.  Note that there is an uncertainty of order unity in the above estimate.

\begin{figure}[htbp]
\begin{center}
\includegraphics[scale=0.6]{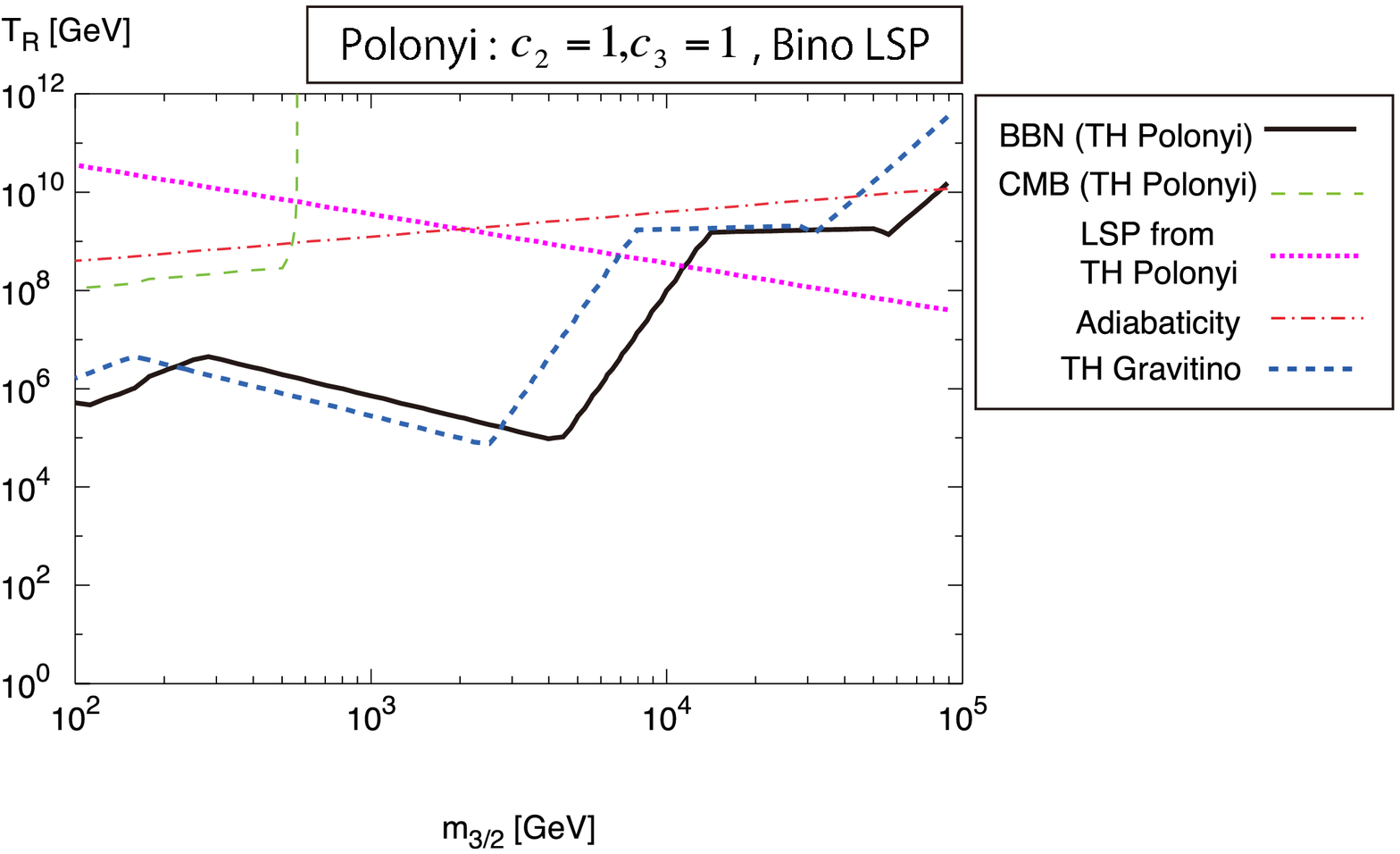}
\vskip 1cm
\includegraphics[scale=0.6]{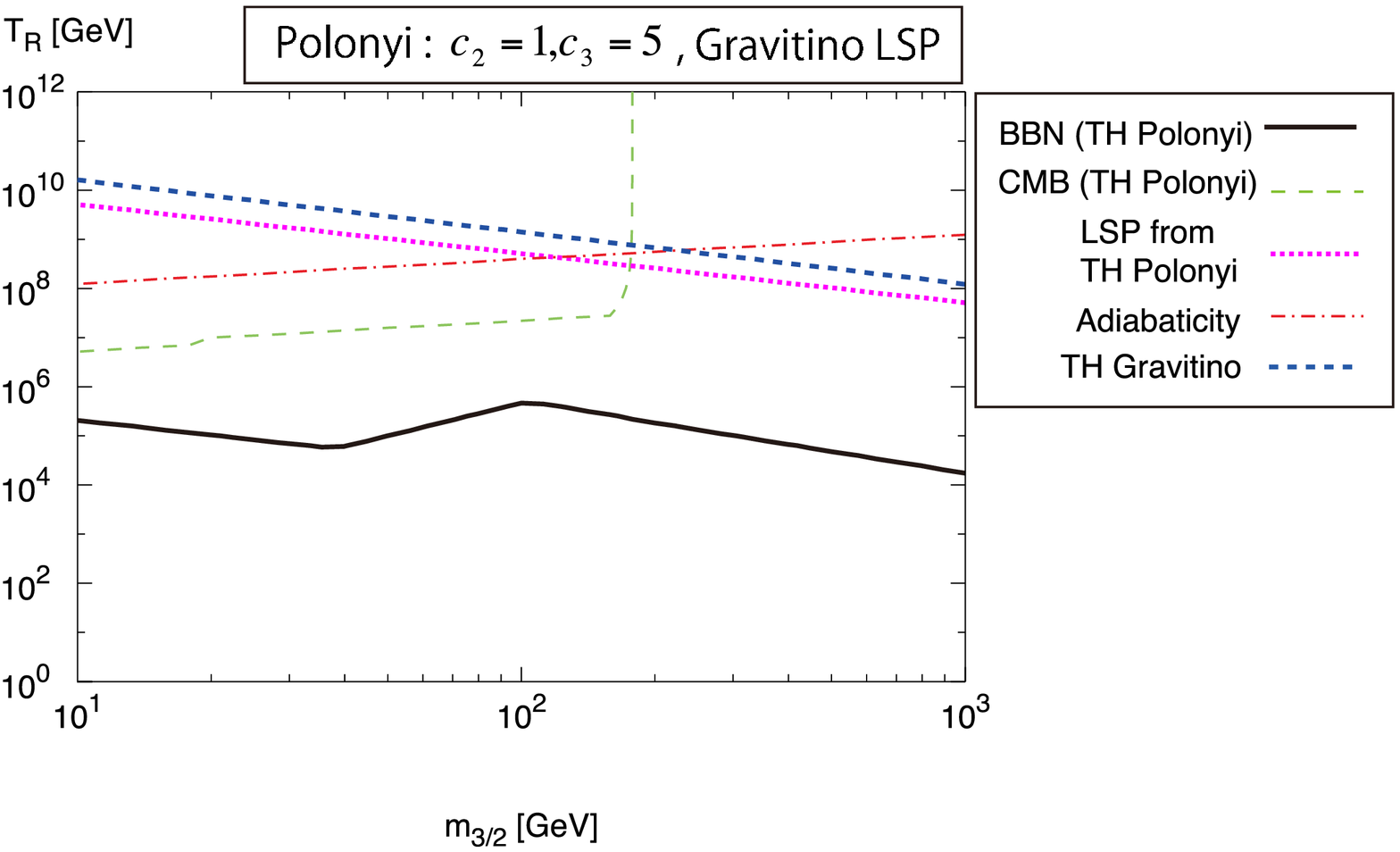}
\caption{
	Upper bounds on the reheating temperature for the Polonyi model with $m_z = m_{3/2}$
	$(c_2=1)$
	for $c_3=1$ (top) and $c_3=5$ (bottom). In the top (bottom) panel, LSP is the bino/higgsino (gravitino).
	 ``BBN (TH Polonyi)'' and ``CMB (TH Polonyi)'' refers to the BBN and CMB 
 	bounds on the thermally produced Polonyi,
	``LSP from TH Polonyi'' to the LSP overproduction bound from the Polonyi decay,
	``TH Gravitino'' to the bound from the gravitino thermal production,
	taking account of the gravitino decay effects on BBN and LSP overproduction,
	and ``Adiabaticity'' to the bound for successful adiabatic suppression on the 
	coherent oscillation of the Polonyi.
	Note that in this figure we do not take into account the thermal relic abundance of the bino/higgsino.
 }
\label{fig:Polonyi}
\end{center}
\end{figure}

Using the Polonyi and gravitino abundances, we can derive cosmological constraints
on the reheating temperature. We show  in Fig.~\ref{fig:Polonyi} upper bounds on the reheating temperature as a function of the gravitino mass.
Here we set $m_z=m_{3/2}$,   $c_3=1(5)$ and $c_5 = 1$ in the top (bottom) panel.
The precise value of $c_4$ is not relevant
for the following arguments as long as $|c_4| \lesssim {\cal O}(1)$.
Here and in what follows the GUT relation among gaugino masses is assumed, unless otherwise stated. 
In the top (bottom) panel, the LSP is the bino- or higgsino-like neutralino (gravitino). 
The constraints on the thermal relic abundance of the SSM LSP are not
taken into account in the both panels, which will be discussed shortly. 
The meaning of each line is as follows.
 ``BBN (TH Polonyi)'' and ``CMB (TH Polonyi)'' refers to the BBN~\cite{Kawasaki:2004qu,Jedamzik:2006xz,Cyburt:2010vz} 
 and CMB~\cite{Fixsen:1996nj} bounds on the thermally produced Polonyi,
``LSP from TH Polonyi'' to the LSP overproduction bound from the Polonyi decay,\footnote{
	Note that in this figure we do not take into account the thermal relic abundance of the LSP.
	In order to avoid the overabundance, the LSP mass cannot be arbitrarily large,
	although the precise thermal abundance depends on other SUSY particle masses.
}
``TH Gravitino'' to the bound from the gravitino thermal production,
taking account of the gravitino decay effects on BBN and LSP overproduction,
and ``Adiabaticity'' to the bound for successful adiabatic suppression on the 
coherent oscillation of the Polonyi.
We can see from Fig.~\ref{fig:Polonyi}  
that the Polonyi problem can be solved for a sufficiently low reheating temperature, $T_{\rm R} \lesssim \GEV{5}$.
The adiabatic solution is therefore an attractive solution to the Polonyi problem.  
On the other hand, the reheating temperature above $\GEV{9}$ is not allowed, and therefore
the thermal leptogenesis does not work~\cite{Fukugita:1986hr}, if all the constraints are taken at face value. 

%%%%%%%%%%%%%%%
%%%%%%%%%%%%%%%

Now let us discuss if some of the constraints could be relaxed.
In the case of the bino/higgsino LSP (top panel), the BBN constraint on the gravitino decay is so stringent 
for $m_{3/2} \lesssim 10{\rm \,TeV}$. So let us focus on the heavy gravitino of mass $\gtrsim 10$~TeV,
where the most stringent constraint  comes from the LSP overabundance (dotted pink).  
In order to weaken the bound, we need to suppress both the thermal and non-thermal relic of the LSP.
(Note that the thermal relic abundance of the LSP is not taken into account in the figure.)
In order to suppress the thermal relic abundance, the higgsino component should be sizable.
The non-thermal component can be suppressed 
if the LSP mass is much lighter than the gravitino.
Thus, one possibility to relax the constraint in the heavy gravitino region 
is to assume that the LSP is the light mixed bino-higgisino neutralino, namely, 
$m_{\rm bino} \sim \mu \ll m_{3/2}$, which requires $c_3 \ll 1$.
Then the SUSY mass spectrum would be similar to the focus point region~\cite{Feng:1999mn}, 
which is usually considered as a region where
thermal leptogenesis works when only the gravitino constraint is considered.  
The bound can be similarly relaxed for the (purely) higgsino-like
LSP,  much lighter than the SSM gauginos and sfermions.
In both cases the bound can be relaxed by an order of several tens to hundred, and the reheating
temperature can be as high as $\oGEV{9}$.
Another possibility to relax the bound is to introduce R-parity violation; then the constraint disappears, but another
DM candidate such as an axion would be needed.

In the case of the gravitino LSP (bottom panel),  the BBN constraint
on the Polonyi decay is so stringent that it is difficult to make the thermal leptogenesis viable. 
This should be contrasted to 
the usual case in which the reheating temperature can be as high as $\oGEV{9}$ at $m_{3/2} = \oGEV{2}$
in the absence of the Polonyi field.

In summary, the adiabatic solution, with fine-tuning on the Polonyi potential at 0.1\% level, works in the minimal set-up 
if $T_{\rm R}$ is sufficiently low, $T_{\rm R} \lesssim \GEV{5}$. 
The constraint on the reheating temperature can be relaxed in several ways.  In the case of the neutralino LSP,
the thermal letpogenesis becomes possible for the gravitino mass as heavy as ${\cal O}(10)$\,TeV, 
if the lightest neutralino is
the mixed bino-higgsino or purely higgsino-like LSP, or if the R-parity is violated.
In the case of the mixed bino-higgsino LSP, the SUSY mass spectrum is similar to the focus point region. 
On the other hand, the thermal leptogenesis~\cite{Fukugita:1986hr} does not work 
in the case of the gravitino LSP because of the stringent BBN bound on the Polonyi decay.

%%%%%%%%%%%%%%%%%%%%%%%%%%%%%%%%%%%%
\subsection{Extended adiabatic solution}
%%%%%%%%%%%%%%%%%%%%%%%%%%%%%%%%%%%%

Now let us extend the minimal set-up to allow other parameters to be enhanced. 
In particular we focus on the case of $c_1\gg 1$ and $c_2\gg 1$. Such an enhancement
of $c_2$ is indeed expected in a certain theoretical framework.
In this case the Polonyi is much heavier than the gravitino and 
the cosmological problems associated with the thermally produced Polonyi can be relaxed.
%On the other hand, the non-thermal gravitino production from the inflaton decay may also be enhanced,
%which  cause a cosmological problem in some inflation models. We will discuss this issue in the next subsection.

In addition to $c_1$ and $c_2$, the Polonyi may also couple to SSM fields with enhanced interactions.
Actually, as shown in Ref.~\cite{Takahashi:2010uw}, it is conceivable that all the Polonyi couplings are
universally enhanced, $c_1,c_2,c_3,c_4, c_5 = {\mathcal O}(10)$. The gaugino mass is
determined by $c_3$, while the sfermion mass depends on $c_4$ and $c_5$. If $c_3$, $c_4$, and/or
$c_5$ are larger than $\oone$, the gauginos and sfermions are heavier than the gravitino. 
On the other hand, the Polonyi abundance (\ref{modTP}) depends on $c_3$ and $c_5$.
In the following we vary those parameters and analyze how they affect the cosmological bounds.

The Polonyi is much heavier than the gravitino, if $c_2 \gg 1$.
Then the Polonyi decays into a pair of the gravitinos,  as well as into the gauge bosons.\footnote{
	Decay into gauginos are less efficient for $m_{3/2} \ll m_z$~\cite{Endo:2006ix}.
}
The decay rate into the gravitino pair is given by
\begin{equation}
	\Gamma_{Z\to 2\psi_{3/2}} = \frac{1}{96\pi}\frac{m_z^5}{m_{3/2}^2M_P^2}.
\end{equation}
In fact, $Z\to 2\psi_{3/2}$ is often the dominant decay mode. If this is the case, the branching fraction of decay into visible particles is suppressed,
which relaxes the BBN constraint on the Polonyi decay if the gravitino is the LSP and stable.
%Instead, the constraint that gravitinos produced non-thermally by the Polonyi decay 
%must not exceed the DM abundance sets an upper bound on the reheating temperature.

The Polonyi and gravitino abundances, given by Eqs.~(\ref{modTP}) and (\ref{gravTP}), respectively,
are enhanced for $c_3 \gg 1$, namely, $m_{\tilde g} \gg m_{3/2}$. Although the partial decay rate of the Polonyi field
into gauge bosons is enhanced for $c_3 \gg 1$, the branching fraction remains small if the gravitino production
is the main decay mode. 

\begin{figure}[ht!]
\begin{center}
\includegraphics[scale=0.6]{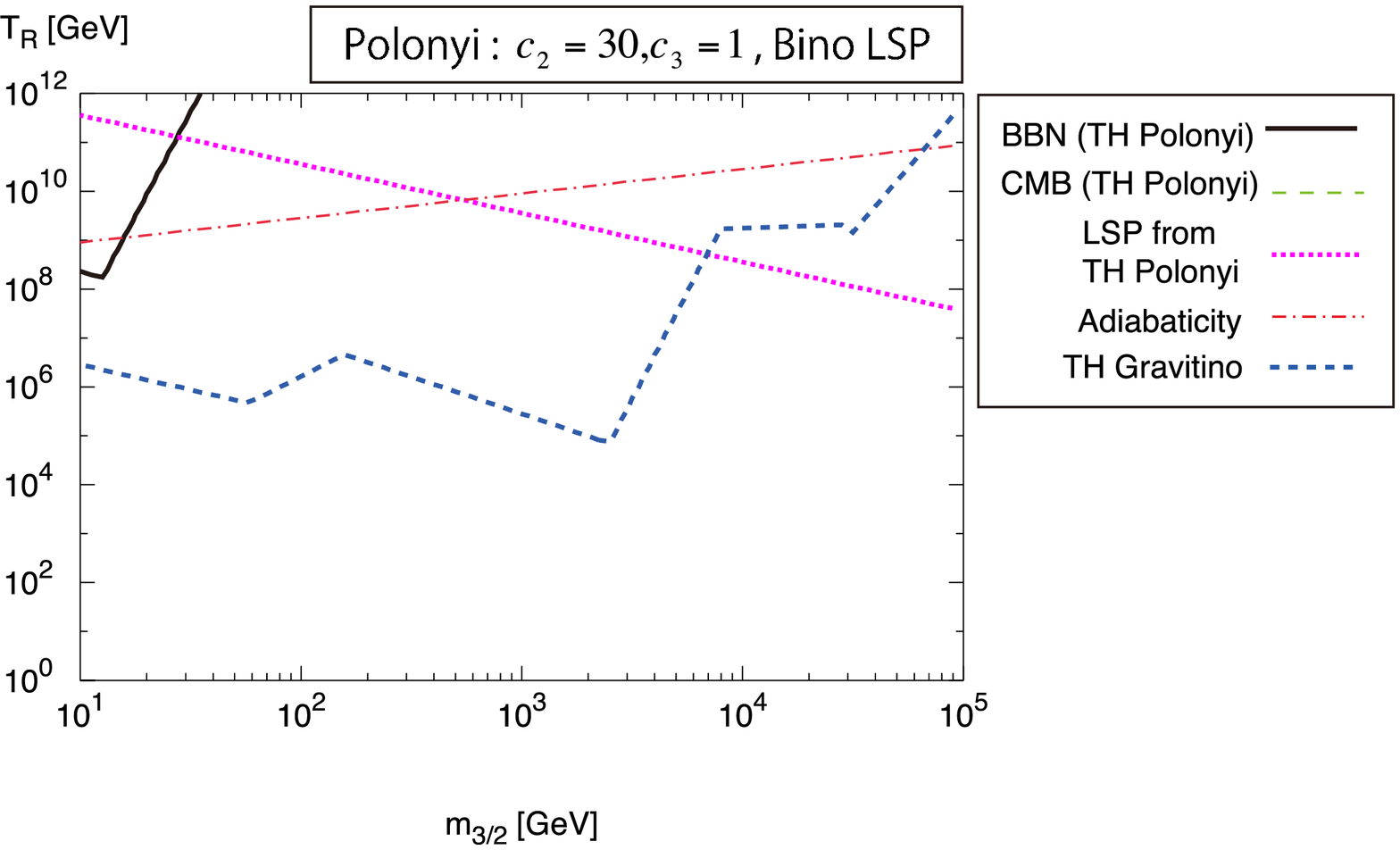}
\vskip 1cm
\includegraphics[scale=0.6]{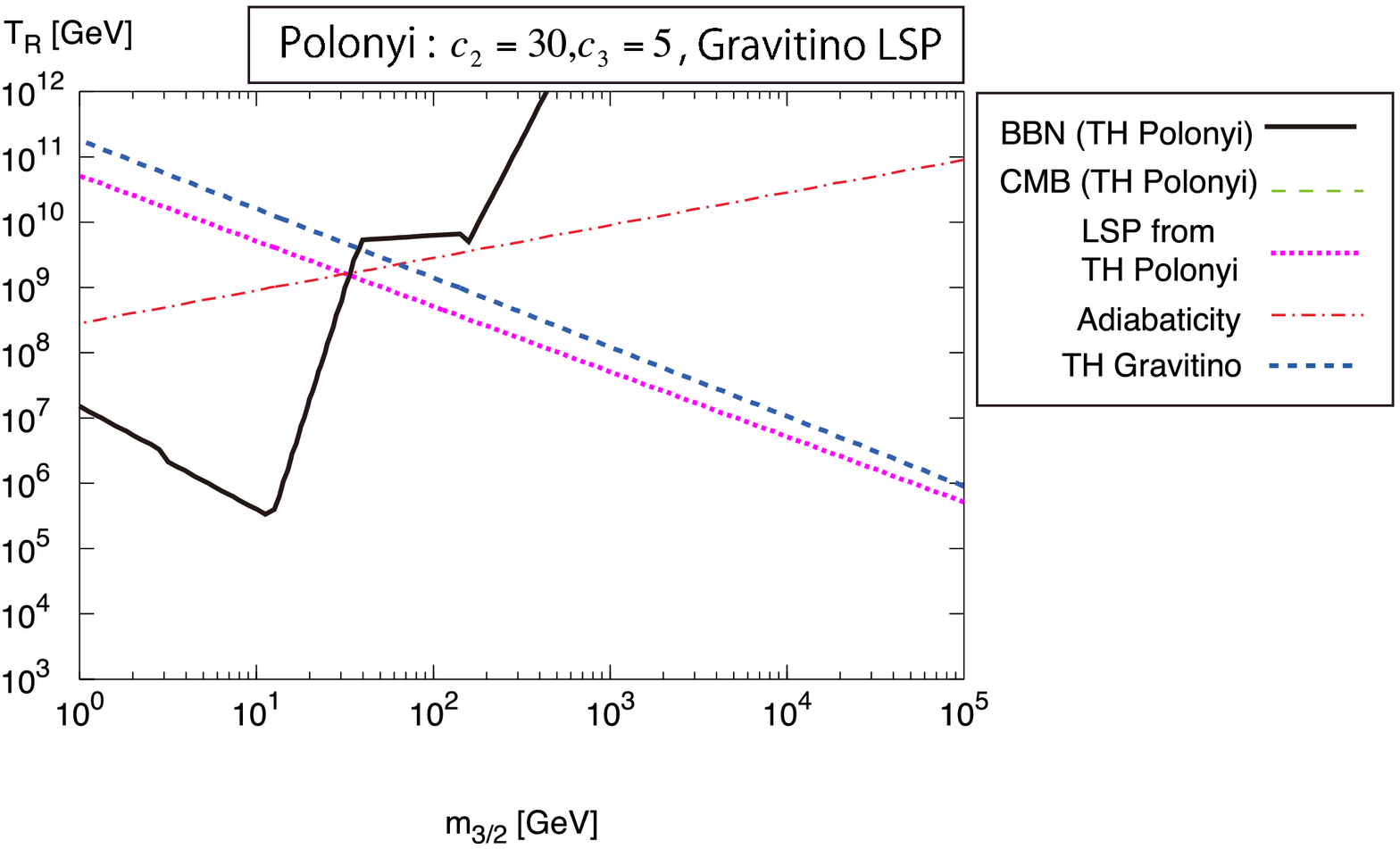}
\caption{
	Same as Fig.~\ref{fig:Polonyi} but for 
	$c_2=30$ and $c_3=1$ (top) and
	$c_2=30$ and $c_3=5$ (bottom) with $c_3=c_5$.
	In the top (bottom) panel, LSP is the bino (gravitino). 
 }
\label{fig:c2=20}
\end{center}
\end{figure}

The upper bounds on the reheating temperature are shown for $c_2=30$ and $c_3=c_5=1(5)$
in the top(bottom) panel in Fig.~\ref{fig:c2=20}.  The meaning of each line is same as  in Fig.~\ref{fig:Polonyi}.
In the top panel, the LSP is either bino- or higgsino-like (or mixed bino-higgsino) neutralino,
while the gravitino is the LSP in the bottom panel. 
Compared to Fig.~\ref{fig:Polonyi}, the gravitino constraint is same as before in both cases, 
while the BBN constraint on the Polonyi decay is significantly relaxed.
This is because the Polonyi mass is much heavier  than the gravitino and it decays
before BBN begins, or   at an early stage of BBN where the constraint from 
the helium overproduction is relatively weak. 

In the top panel, the overall constraints are similar to Fig.~\ref{fig:Polonyi},
and the constraint from the LSP overabundance in the heavy gravitino region can be relaxed 
in a similar way to the previous case.
The reheating temperature as high as $10^9$~GeV is allowed only 
for the heavy gravitino case of $m_{3/2}\sim 10$\,TeV.
Even if we demand a moderate reheating temperature of $T_{\rm R} \gtrsim 10^6$~GeV for 
non-thermal leptogenesis~\cite{Asaka:1999jb,Hamaguchi:2001gw},
we need a relatively heavy gravitino, $m_{3/2} \gtrsim 5$~TeV.
On the other hand, in the bottom panel, since the BBN constraint
on the Polonyi field is significantly relaxed, there appears an interesting parameter region at  $m_{3/2}\sim 100$~GeV
with $T_{\rm R}\sim 10^9$~GeV where the thermal leptogenesis may work successfully.
Here it should be noted that $c_3$ cannot be much larger since it enhances the 
gaugino mass and hence the gravitino thermal production.
If we demand $T_{\rm R}\gtrsim 10^6$~GeV for non-thermal leptogenesis,
a broader region for the gravitno mass $\mathcal O(10){\rm GeV}$-$\mathcal O(10){\rm TeV}$ is allowed.
In this region gravitino is the LSP, while sfermions are  heavier than the gauginos.
This resembles the mass spectrum in the so-called focus-point region~\cite{Feng:1999mn}.

%%%%%%%%%%%%%%%
%%%%%%%%%%%%%%%

Note that the effect of the next-to-lightest SUSY particle (NLSP) is not taken into account in the figure.
In fact, the long-lived NLSP significantly affects BBN in the case of gravitino LSP and hence we need
a small $R$-parity violation in order for the NLSP to decay well before BBN begins.
The NLSP lifetime depends on the pattern of $R$-parity violation and its magnitude.
In particular, if the $R$-parity is violated by the tri-linear interaction such as $u^c d^c d^c$ or $LLe^c$,
the NLSP lifetime depends crucially on the sfermion masses.
The displaced vertex may give useful information on discriminating the model.

%%%%%%%%%%%%%%%%%%%%%%%%%%%%%%%%%%%%
\section{The moduli problem in gauge-mediation}
\label{sec:4}
%%%%%%%%%%%%%%%%%%%%%%%%%%%%%%%%%%%%

The gauge-mediated SUSY breaking (GMSB) models~\cite{Giudice:1998bp} have attractive features
that they do not suffer from the SUSY flavor problem.
In the GMSB model, the Polonyi-like field is not always required in the SUSY breaking sector.
Thus there may not be a Polonyi problem.
But even in this case there may exist modulus fields especially if
the four dimensional supergravity arises from the compactification of the extra dimensions in string theory.
These moduli are expected to have masses of the gravitino and Planck-suppressed interactions, 
and hence they cause serious cosmological problems.

In this case the relevant parameters are the inflaton-modulus coupling $c_1$, which must be enhanced in order to 
solve the moduli problem.
As shown in Fig.~\ref{fig:tuning}, the moduli oscillation can be sufficiently suppressed by the adiabatic solution without fine-tuning 
for $m_z \lesssim 1$\,MeV. 
However, we still need to care about thermally produced moduli.
In the GMSB model, the moduli, as well as the gravitino, are light and the lifetime are longer than the case of gravity-mediation.
If the modulus survives after the recombination epoch, the decay produced photons contribute to the
diffuse X($\gamma$)-ray background~\cite{Sreekumar:1997un}
which may easily exceed the observational limit~\cite{Kawasaki:1997ah,Kawasaki:2007mk}.

\begin{figure}[htbp]
\begin{center}
\includegraphics[scale=0.6]{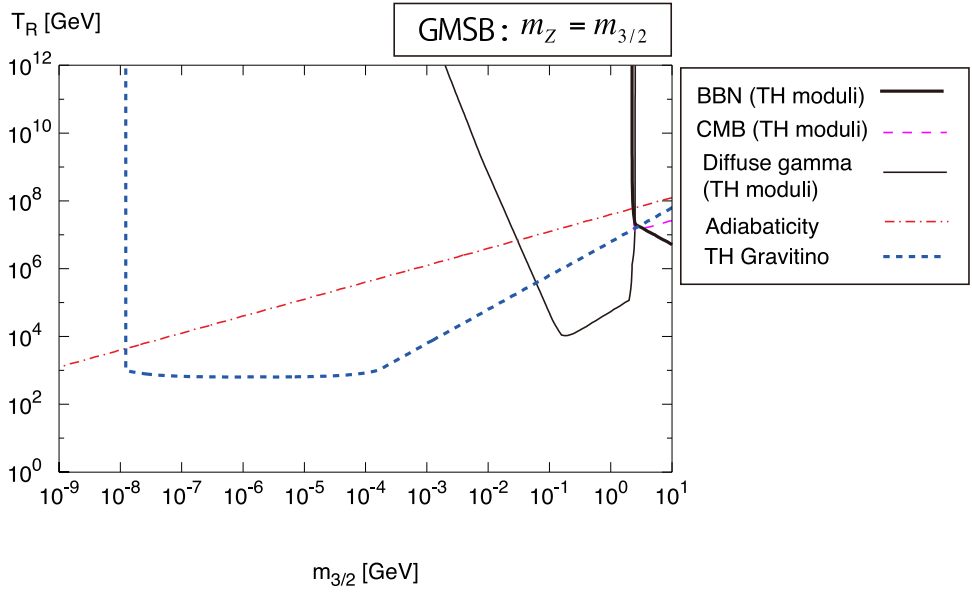}
\vskip 1cm
\includegraphics[scale=0.6]{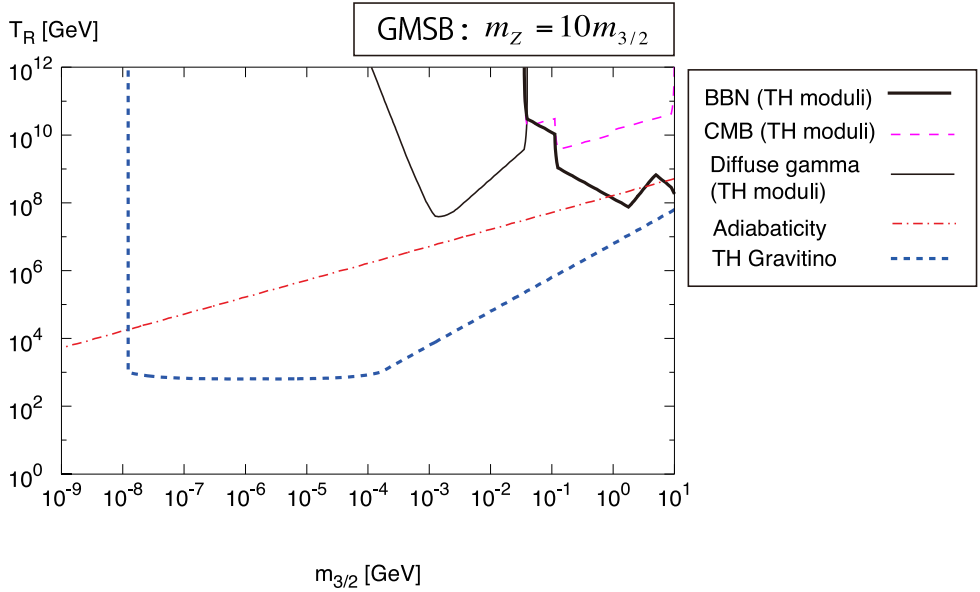}
\caption{
	Upper bounds on the reheating temperature in GMSB models with moduli.
	We take $m_z = m_{3/2}$ (top) and $m_z = 10m_{3/2}$ (bottom) and $m_{\tilde g}$=1~TeV.
	``BBN (TH moduli)'', ``CMB (TH moduli)'' and ``Diffuse gamma (TH moduli)''
	refer to the BBN, CMB and diffuse gamma-ray flux bounds on the thermally produced moduli,
	%``LSP from TH moduli'' to the LSP overproduction bound from the decay of thermally produced moduli,
	``Adiabaticity'' to the bound in order for suppress the Polonyi abundance,
	and ``TH Gravitino'' to the usual bound from the gravitino thermal production.
 }
\label{fig:GMSB}
\end{center}
\end{figure}

Fig.~\ref{fig:GMSB} shows constraints on the reheating temperature in the GMSB model with moduli.
We take $m_z = m_{3/2}$ (top) and $m_z = 10m_{3/2}$ (bottom) and $m_{\tilde g}$=1~TeV.
``BBN (TH moduli)'', ``CMB (TH moduli)'' and ``Diffuse gamma (TH moduli)''
refer to the BBN, CMB and diffuse gamma-ray flux bounds on the thermally produced moduli,
``Adiabaticity'' to the bound in order for suppress the Polonyi abundance and
``TH Gravitino'' to the usual bound from the gravitino thermal production.

First it is remarkable that the ultra-light gravitino mass region, $m_{3/2}\lesssim 16$~eV, where 
it was believed that there is no upper bound on the reheating temperature,
is clearly inconsistent with high-reheating temperature scenario
once we demand that the adiabatic mechanism suppresses the modulus abundance.\footnote{
	The smallness of the modulus abundance in this mass range may be explained 
	by the anthropic arguments without invoking the adiabatic suppression, 
	since in this region the constraint comes from the modulus 
	overabundance as the DM.
}

For the case of intermediate gravitino mass of $\mathcal O(10)$\,GeV,
the situation is similar to that studied in the previous section.
It may be consistent with non-thermal leptogenesis scenario 
if the modulus mass is enhanced compared to the gravitino mass
under the broken $R$-parity.

%%%%%%%%%%%%%%%
%%%%%%%%%%%%%%%

%%%%%%%%%%%%%%%%%%%%%%%%%%%%%%%%%%%%
\section{The moduli problem in anomaly-mediation}
\label{sec:5}
%%%%%%%%%%%%%%%%%%%%%%%%%%%%%%%%%%%%

Finally we mention the moduli problem in the anomaly-mediated SUSY breaking 
(AMSB) model~\cite{Randall:1998uk}.
As noted in Introduction, the moduli problem is milder than that in the case of
gravity- or gauge-mediation since the moduli are heavy enough to decay before BBN.
But it was recognized in Ref.~\cite{Endo:2006zj} that the gravitinos produced by the modulus decay
cause cosmological problems.
Even if the gravitinos are also heavy enough, the abundance of
LSPs produced by the modulus/gravitino decay is too much.
A possible solution is to introduce a small $R$-parity violation in order for the LSP to decay.
In any case, the modulus decay produces huge amount of entropy and the preexisting
baryon asymmetry is significantly diluted.\footnote{
	The amount of dilution factor is roughly given by $\sim T_z/T_{\rm R}$ where $T_z$ is the modulus decay temperature.
	The most baryogenesis models do not work under this condition, except for some parameter ranges in the
	Affleck-Dine mechanism~\cite{Kawasaki:2007yy}. 
}

\begin{figure}[htbp]
\begin{center}
\includegraphics[scale=0.6]{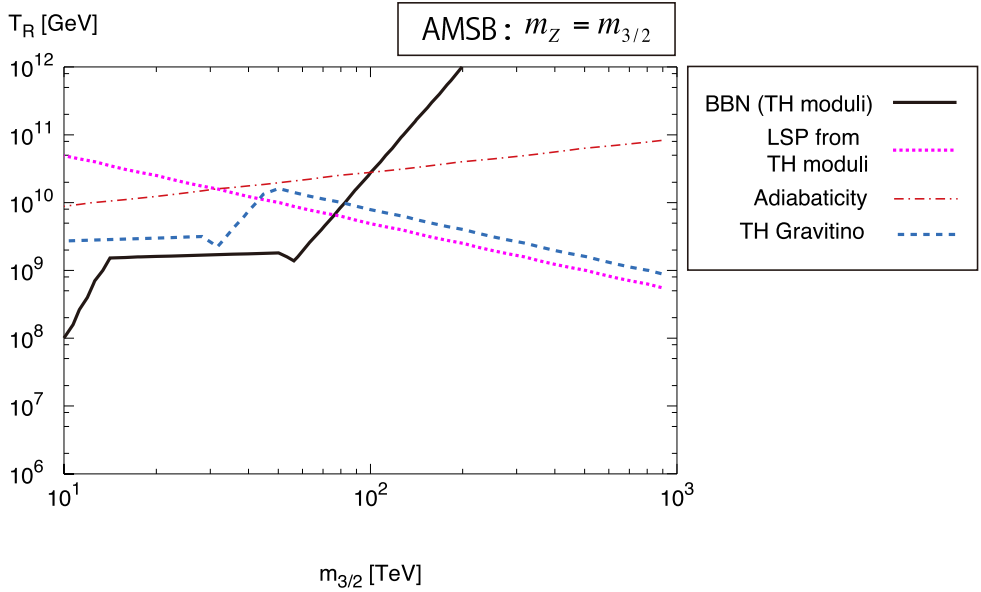}
\vskip 1cm
\includegraphics[scale=0.6]{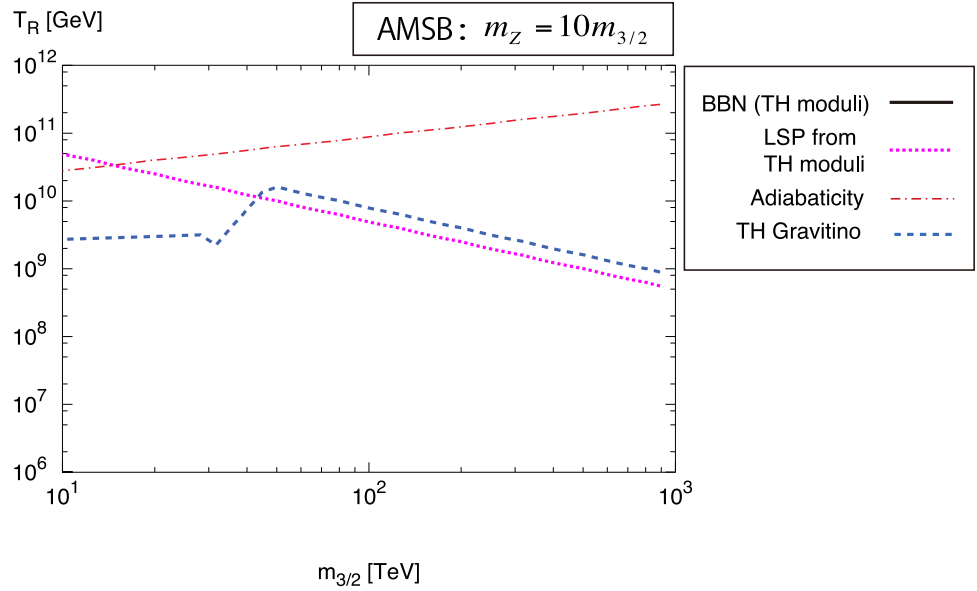}
\caption{
	Upper bounds on the reheating temperature in AMSB models with moduli.
	We take $m_z = m_{3/2}$ (top) and $m_z = 10m_{3/2}$ (bottom).
	Meaning of each lines is same as those in Fig.~\ref{fig:GMSB},
	except for the ``LSP from TH moduli'' which denotes the LSP overproduction bound from
	the decay of thermally produced moduli.
 }
\label{fig:AMSB}
\end{center}
\end{figure}

Here we consider the adiabatic solution to the moduli problem in the AMSB model,
which is free from the problems of late-time entropy production and hence the dilution of the baryon asymmetry.
As shown in Fig.~\ref{fig:tuning}, we need a tuning on the modulus potential at the level of 
$\mathcal O(10^{-3})$ or $\mathcal O(10^{-2})$
in order to solve the moduli problem in the original sense for a relatively high-reheating temperature
consistent with thermal or nonthermal leptogenesis.
Again, however, we still need to take account of thermal production of the moduli.
Fig.~\ref{fig:AMSB} shows upper bounds on the reheating temperature in AMSB models with moduli.
We take $m_z = m_{3/2}$ (top) and $m_z = 10m_{3/2}$ (bottom).
The meaning of each line is same as those in Fig.~\ref{fig:GMSB},
except for the ``LSP from TH moduli'' which denotes the LSP overproduction bound from
the decay of thermally produced moduli.
Note that in this case the ``TH gravitino'' bound includes the LSP overproduction from the gravitino decay.
It is found that the gravitino mass of around $100{\rm TeV}$ is compatible with the
thermal leptogenesis scenario ($T_{\rm R} \gtrsim 10^9$\,GeV).
We do not need an $R$-parity violation in this case.

Notice that the Wino is the LSP in AMSB models and its mass is given by
$m_{\tilde W} =(g_2^2/16\pi^2)m_{3/2}\sim 2.6\times 10^{-3}m_{3/2}$.
In the parameter region considered above, 
the Wino produced non-thermally by the gravitino/modulus decay can be the dominant component of DM.
But the Wino has rather large annihilation cross section and
the Wino DM mass is limited from the observation 
of light element abundances~\cite{Jedamzik:2004ip,Hisano:2008ti},
cosmic microwave background anisotropy~\cite{Galli:2009zc,Kanzaki:2009hf}
and gamma-rays by the Fermi satellite~\cite{Abdo:2010ex}.
These constraints demand $m_{\tilde W} \gtrsim 200$\,GeV
which translates into the bound on the gravitino mass as $m_{3/2} \gtrsim 80$\,TeV
if the DM mainly consists of the Wino.
%Notice that a single-field new inflation model, which is favored from the viewpoint of adiabatic suppression 
%on the modulus abundance, is consistent with such a large gravitino mass for a certain parameter range~\cite{Ibe:2006ck}.

%%%%%%%%%%%%%%%
%%%%%%%%%%%%%%%

%%%%%%%%%%%%%%%%%%%%%%%%%%%%%%%%%%%%
\section{Conclusions and discussion}
\label{sec:6}
%%%%%%%%%%%%%%%%%%%%%%%%%%%%%%%%%%%%

One of the major obstacles to construct a consistent cosmological scenarios 
in the most SUSY breaking models is the cosmological Polonyi/moduli problem.
We have examined carefully the adiabatic solution to the cosmological Polonyi/moduli problem,
and found that the Polonyi/moduli oscillation is necessarily induced at the end of inflation, and its
abundance depends on the inflation models.
As a result, the Polonyi/moduli problem cannot be solved without further tuning on the potential parameters
for a broad range of the modulus mass. The possible ranges of the moduli mass 
where no severe tuning is needed 
for solving the moduli problem are either $m_z \lesssim 1$\,MeV or $m_z \gtrsim 10$\,TeV (see Fig.~\ref{fig:tuning}).
In the former case, the reheating temperature cannot be larger than $\sim 1$\,TeV 
from the gravitino constraint (see Fig.~\ref{fig:GMSB}).
In the latter case, the reheating temperature can be as high as $\sim 10^6$\,GeV 
with a mild tuning of $\mathcal O(0.1-0.01)$
and may be consistent with non-thermal leptogenesis scenario~\cite{Asaka:1999jb,Hamaguchi:2001gw}. 
If we allow a tuning at the $1\%$ level, the thermal leptogenesis requires $m_z \gtrsim 100$\,TeV.
On the other hand, for a very low reheating temperature of $\mathcal O(1)$\,MeV, the adiabatic suppression mechanism can
solve the Polonyi/moduli problem without tuning for every range of Polonyi/modulus mass. 

We have also pointed out that Polonyi/moduli production from scatterings in thermal
bath can be relevant, even if the coherent oscillations are suppressed by the adiabatic
mechanism. If the Polonyi/moduli have Planck-suppressed interactions with the SSM
particles, their abundance is comparable to that of the transverse component of the gravitino.
In fact, the cosmological constraint on the Polonyi field produced from thermal scattering 
has turned out to be stringent for a broad range of the gravitino mass (see Fig.~\ref{fig:Polonyi}).

Since the adiabatic solution necessitates the enhanced Polonyi/moduli-inflaton
coupling, other couplings may also be  enhanced, depending on the origin of the enhanced
coupling.  We have studied several cases where the Polonyi self-coupling $(c_2)$
and/or the Polonyi-matter couplings $(c_3, c_4, c_5)$ are enhanced. 
In the gravity-mediation model, the Polonyi field mainly decays into a pair of gravitinos,
if its self-coupling is enhanced. 
This relaxes the cosmological constraint on the Polonyi abundance when the gravitino is the LSP. 
Once we allow a fine-tuning on the Polonyi potential at the level of 
$\mathcal O(10^{-4})$ or $\mathcal O(10^{-3})$ ,
the adiabatic solution works in consistent with
thermal or nonthermal leptogenesis scenario, for some gravitino mass ranges (see Fig.~\ref{fig:tuning}). 
In particular there appears a new interesting parameter space where the gauginos are
relatively light ($\sim \mathcal O(100)$\,GeV) and the gravitino is the LSP.
A small amount of R-parity violation is needed to avoid the BBN constraint on the NLSP decay.
Sfermion masses can be comparable to or much heavier than the gaugino mass depending on the parameters $c_4$ and $c_5$.
We have also commented that non-thermal gravitino production from the inflaton decay may be enhanced in this setup.

In the gauge-mediated SUSY breaking model, there may also exist light moduli, and
we have studied the adiabatic solution to the moduli problem.
It is found that the ultra-light gravitino scenario $(m_{3/2}<16{\rm eV})$,
where the gravitino problem does not exist, is not consistent with 
the adiabatic solution unless $T_{\rm R}\lesssim 1$\,TeV,
although the modulus amplitude may be anthropically tuned without the adiabatic suppression mechanism.

In the anomaly-mediation model, once the modulus oscillation is suppressed 
by the adiabatic suppression mechanism and some tuning,
constraints from thermally produced moduli are rather weak. The reheating temperature can be high enough to
be consistent with (non-)thermal leptogenesis. 

We conclude that there are many non-trivial constraints for the adiabatic suppression on the moduli to work successfully.
Even in the presence of adiabatic suppression, we need a tuning on the modulus potential
in order to avoid the moduli problem especially for a relatively high reheating temperature (see Fig.~\ref{fig:tuning}), 
although the required amount of tuning is significantly relaxed. 
For a low reheating temperature, on the other hand, the adiabatic suppression mechanism can solve the Polonyi/moduli problem
without any fine-tuning.
Thermal production of the moduli also set a severe constraint on the reheating temperature for relatively light moduli 
of $m_z \lesssim 1$\,TeV.
These aspects of the moduli should be taken into account when the solution to the moduli problem 
in the adiabatic suppression is discussed.

%%%%%%%%%%%%%%%%%%%%%%%%%%%%%%%%%%%%
\section*{Acknowledgment}
%%%%%%%%%%%%%%%%%%%%%%%%%%%%%%%%%%%%

This work was supported by the Grant-in-Aid for Scientific Research on
Innovative Areas (No. 21111006) [KN and FT], Scientific Research (A)
(No. 22244030 [KN and FT], 21244033 [FT], 22244021 [TTY]), and JSPS Grant-in-Aid for
Young Scientists (B) (No. 21740160) [FT].  This work was also
supported by World Premier International Center Initiative (WPI
Program), MEXT, Japan.

%%%%%%%%%%%%%%%%%%%%%%%%%%%%%%%%%%%%

%%%%%%%%%%%%%%%%%%%%%%%%%%%%%%%%%%%%

\end{document}